%% file: ms.tex
\documentclass[twocolumn]{aastex631}

\usepackage{subfigure}
\usepackage{url}
\usepackage{hyperref}
\usepackage{epsfig}
\usepackage{graphicx}
\usepackage{sidecap}
\usepackage{hanging}
\usepackage{color}
\usepackage{multirow}
\usepackage{enumerate}
\usepackage{natbib}
\usepackage{amssymb}
\usepackage{amsmath}
\usepackage[bottom]{footmisc}
\usepackage[T1]{fontenc}

\newcommand{\kepler}{{Kepler}}

\providecommand{\e}[1]{\ensuremath{\, \times \, 10^{#1}}}

\newcommand{\koi}{{KOI-289}}
\newcommand{\plb}{{Kepler-511~b}}
\newcommand{\plc}{{Kepler-511~c}}
\newcommand{\host}{{Kepler-511}}

\providecommand{\bjdtdb}{\ensuremath{\rm {BJD_{TDB}}}}

\providecommand{\msun}{\ensuremath{\,M_\Sun}}
\providecommand{\rsun}{\ensuremath{\,R_\Sun}}
\providecommand{\lsun}{\ensuremath{\,L_\Sun}}
\providecommand{\mj}{\ensuremath{\,M_{\rm J}}}
\providecommand{\rj}{\ensuremath{\,R_{\rm J}}}
\providecommand{\me}{\ensuremath{\,M_{\rm E}}}
\providecommand{\re}{\ensuremath{\,R_{\rm E}}}

\shorttitle{GOT `EM V. The Giant in Kepler-511 that Ran Away}
\shortauthors{Y. Chachan et al.}

\begin{document}

\title{Giant Outer Transiting Exoplanet Mass (GOT `EM) Survey. V. \\ Two Giant Planets in Kepler-511 but Only One Ran Away}

\correspondingauthor{Yayaati Chachan}
\email{ychachan@ucsc.edu}


\author[0000-0003-1728-8269]{Yayaati Chachan}
\affiliation{Department of Astronomy \& Astrophysics, University of California, Santa Cruz, CA 95064, USA}
\affiliation{Department of Physics and Trottier Space Institute, McGill University, 3600 rue University, H3A 2T8 Montr\'eal, QC, Canada}
\affiliation{Trottier Institute for Research on Exoplanets (iREx), Universit\'e de Montr\'eal, Quebec, Canada}

\author[0000-0002-4297-5506]{Paul A.\ Dalba}
\affiliation{Department of Astronomy \& Astrophysics, University of California, Santa Cruz, CA 95064, USA}

\author[0000-0002-5113-8558]{Daniel P.\ Thorngren}
\affiliation{Department of Physics \& Astronomy, Johns Hopkins University, Baltimore, MD 21210, USA}

\author[0000-0002-7084-0529]{Stephen R.\ Kane}
\affiliation{Department of Earth and Planetary Sciences, University of California Riverside, 900 University Ave, Riverside, CA 92521, USA}

\author[0000-0002-0531-1073]{Howard Isaacson}
\affiliation{{Department of Astronomy,  University of California Berkeley, Berkeley CA 94720, USA}}
\affiliation{Centre for Astrophysics, University of Southern Queensland, Toowoomba, QLD, Australia}

\author[0000-0002-1228-9820]{Eve J. Lee}
\affiliation{Department of Physics and Trottier Space Institute, McGill University, 3600 rue University, H3A 2T8 Montr\'eal, QC, Canada}
\affiliation{Trottier Institute for Research on Exoplanets (iREx), Universit\'e de Montr\'eal, Quebec, Canada}
\affiliation{Department of Astronomy \& Astrophysics, University of California, San Diego, La Jolla, CA 92093, USA}

\author[0000-0002-2949-2163]{Edward W.\ Schwieterman}
\affiliation{Department of Earth and Planetary Sciences, University of California Riverside, 900 University Ave, Riverside, CA 92521, USA}

\author[0000-0001-8638-0320]{Andrew W.\ Howard}
\affiliation{Department of Astronomy, California Institute of Technology, Pasadena, CA 91125, USA}

\author[0000-0001-5133-6303]{Matthew J.\ Payne}
\affiliation{Harvard-Smithsonian Center for Astrophysics, 60 Garden St., MS 51, Cambridge, MA 02138, USA}


\begin{abstract}
Systems hosting multiple giant planets are important laboratories for understanding planetary formation and migration processes. We present a nearly decade-long Doppler spectroscopy campaign from the HIRES instrument on the Keck-I telescope to characterize the two transiting giant planets orbiting Kepler-511 on orbits of 27 days and 297 days. The radial velocity measurements yield precise masses for both planets: $0.100^{+0.036}_{-0.039}$ ($2.6 \sigma$) and $0.44^{+0.11}_{-0.12}$ (4$\sigma$) Jupiter masses respectively. We use these masses to infer their bulk metallicities (i.e., metal mass fraction $0.87 \pm 0.03$ and $0.22 \pm 0.04$ respectively). Strikingly, both planets contain approximately $25-30$ Earth masses of heavy elements but have very different amounts of hydrogen and helium. Envelope mass loss cannot account for this difference due to the relatively large orbital distance and mass of the inner planet. We conclude that the outer planet underwent runaway gas accretion while the inner planet did not. This bifurcation in accretion histories is likely a result of the accretion of gas with very different metallicities by the two planets or the late formation of the inner planet from a merger of sub-Neptunes. Kepler-511 uniquely demonstrates how giant planet formation can produce dramatically different outcomes even for planets in the same system.

\end{abstract}


\section{Introduction} \label{sec:intro}

One of the many interesting findings to come from the discovery and characterization of exoplanet systems in recent years is that multiplanet systems are common. This conclusion about planet multiplicity has been reached by transit and radial velocity (RV) alike \citep[e.g.,][]{Thompson2018, Bryan2019, Rosenthal2021, Zhu2022}. Of multiplanet systems, those with multiple giants pose interesting questions about planetary formation and the evolution of the system as a whole. Each giant planet formed from the same protoplanetary disk, but a variety of initial conditions \citep[e.g.,][]{Miguel2011, Oberg2011}, migration channels \citep[e.g.,][]{Goldreich1980, Rasio1996, Wu2003, Lithwick2011b, Winn2011}, and other potentially stochastic processes \citep[e.g.,][]{Carrera2019} eventually produce the planetary system as it is observed today. By measuring the current orbital and physical properties of giant planets in multiplanet systems, we aim to disentangle these processes and better understand planet formation as a whole.

Multiplanet systems containing transiting exoplanets are a particularly interesting subset because they allow for the measurement of at least one of the planet's radius and thereby its bulk density. In the specific case of a giant exoplanet ($\gtrsim 4$ R$_\oplus$), the bulk composition can lend clues about the planet's heavy element abundance \citep[e.g.,][]{Guillot2006}, which in turn informs its formation and evolution \citep[e.g.,][]{Pollack1996,Mousis2009,Thorngren2016,Hasegawa2018}. Having this constraint for \emph{all} of the known giant planets in a particular system is especially valuable as a tool for investigating different accretion process ongoing in the same system. According to the NASA Exoplanet Archive \citep{Akeson2013}, of the nearly 900 known multiplanet systems, well over 100 have at least one transiting giant planet but only a few dozen contain multiple giant transiting planets\footnote{\url{https://exoplanetarchive.ipac.caltech.edu}}. 

Recent simulations aimed at capturing the impact of disk gap formation on multiple forming giants planets have found that the diverse distribution of mass ratios between giant planets in the same system necessitates different accretion start times as well as some amount of truncation of runaway accretion by disk dispersal \citep{BergezCasalou2023}. Runaway accretion can occur at different times in different locations with a protostellar disk owing to different disk conditions such as disk opacity, aspect ratio, surface density, etc. \citep[e.g.,][]{Ikoma2001, Bitsch2018, Chachan2021}. In this way, mass measurements of multiple giant planets in the same system provide insight into the properties of the local disk in which they both formed. 

Moreover, the final orbital distances of giant planets in multiplanet systems hold clues as to how the planets migrated and initial disk properties, which is especially interesting for comparison with our own multi-giant-planet system. \citet{Griveaud2023} found that pairs of giant planets only migrate inward in low viscosity disks and end up as ``warm Jupiters,'' in contrast to Solar System based theories that suggest that Jupiter and Saturn also migrated outward together \citep[e.g.,][]{Tsiganis2005,Walsh2011b}. 

In this work, we confirm and characterize the \host\ system (\koi), which was previously known to contain two statistically validated giant planets \citep{Morton2016,Valizadegan2022}. Despite the fact that \plb\ and \plc\ have long orbital periods relative to other transiting exoplanets (297 days and 27 days, respectively), both planets transit their host star. Moreover, the factor of 11 difference in their orbital periods (5x in semimajor axis), is much larger than nearly all other well characterized systems of multiple transiting giant exoplanets. 

This work adds another chapter to the Giant Outer Transiting Exoplanet Mass (GOT `EM) Survey, which aims to measure the masses and radii of giant exoplanets on relatively long-period orbits \citep{Dalba2021a,Dalba2021c,Mann2023,Dalba2024}. Well characterized giant planets with orbital periods of 100 days or longer are intrinsically rare owing to their low transit probabilities and the practical difficulty in measuring their masses via Doppler spectroscopy. The GOT `EM survey aims to increase the sample size of planets in this parameter space in support of giant planet formation and evolution theories. To date, this survey has measured the masses and orbital ephemerides of nearly a dozen giant planets with orbital periods between 100 and 1000 days.

This paper is organized as follows. In Section~\ref{sec:obs}, we describe transit observations of \host\ from the Kepler spacecraft and follow-up spectroscopy from Keck Observatory. In Section~\ref{sec:model}, we we model the stellar and planetary properties of this system with \textsf{EXOFASTv2} \citep{Eastman2019} and \textsf{exoplanet} \citep{ForemanMackey2021}. In Section~\ref{sec:res} we present the results of the modeling and take a step further to estimate bulk heavy element abundance of both planets. Interestingly, both planets have similar masses of heavy elements but their bulk metallicity is notably different. In Section~\ref{sec:disc}, we propose theories for how \plb\ and \plc\ formed given their bulk metallicities. Finally, in Section~\ref{sec:conc}, we summarize our analysis and primary findings.


\section{Observations}\label{sec:obs}

In the following sections, we describe how all photometric and spectroscopic observations were collected and processed.


\subsection{Photometric data from Kepler}\label{sec:kepler}

\begin{figure*}
    \centering
    \includegraphics[width=\textwidth]{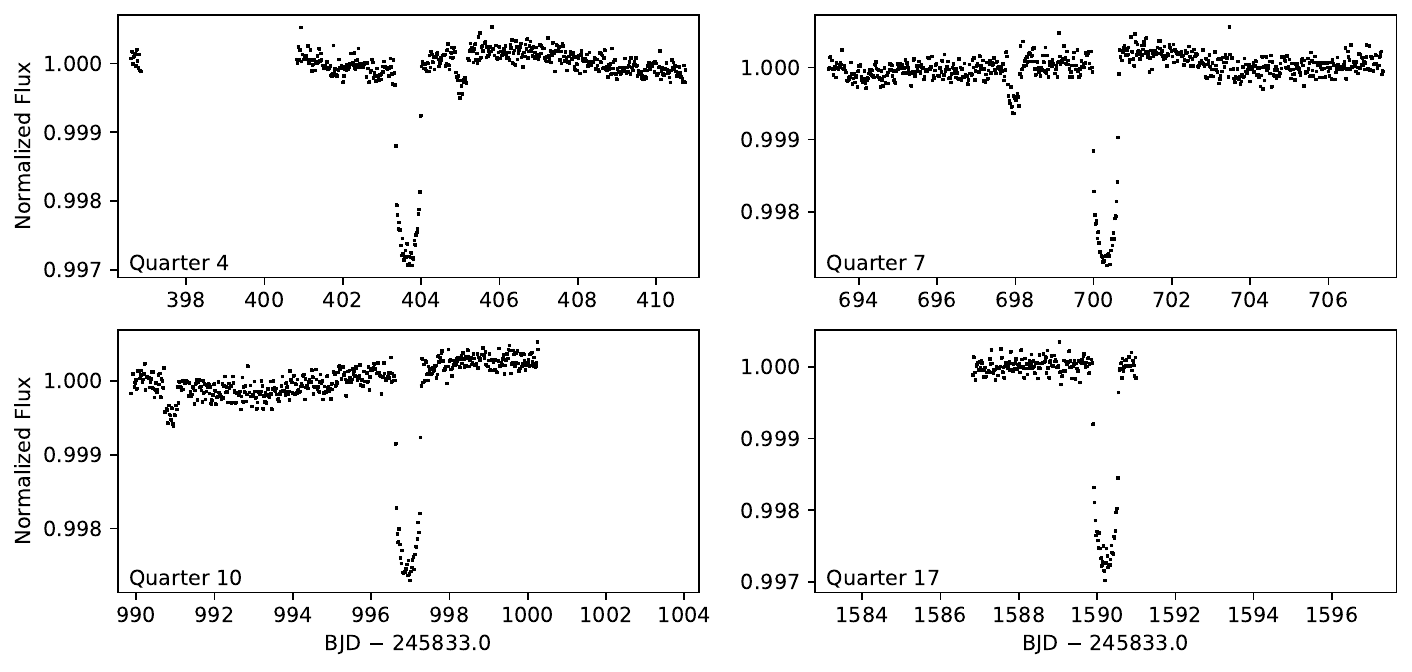}
    \caption{Raw Kepler photometry showing the four transits of \plb\ and three of the transits of \plc. Each panel is centered on the mid-transit time of \plb\ and includes 10 transit durations worth of data before and after. Some baseline data were lost in all quarters except for Quarter 7, and the Quarter 14 transit was lost altogether to a data gap. Even in the raw light curves, the transits of both planets are visible. The stellar variability, which was removed prior to fitting the transit, can be seen as a gentle, low frequency variation in flux.}
    \label{fig:nondetrended_transits}
\end{figure*}

\host\ was observed by the Kepler spacecraft \citep{Borucki2010} in Quarters 0--17 (02-May-2009 to 11-May-2013). We downloaded the data from the Mikulski Archive for Space Telescopes using the \textsf{lightkurve} package \citep{Lightkurve2018}. Most of the data were acquired at long cadence (30~minutes), although some short cadence (1~minute) data were available in Quarters 6 and 7. When both types of data were available, we defaulted to using the short cadence. 

We made use of the Pre-search Data Conditioning Simple Aperture Photometry \citep[PDCSAP;][]{Jenkins2010a,Smith2012,Stumpe2012} which was detrended for various systematic noise sources including dilution and the fraction of \host's flux that is captured by the photometric aperture. The point spread function of \host\ was approximately 1 pixel in radius, and the photometric apertures were approximately 2--3 pixels in radius. Each Kepler pixel subtends $\sim$4~arcseconds. The dilution correction did not account for the nearby source at 3\farcs26 identified by later direct imaging (see Section~\ref{sec:companion}). However, this source is approximately 7 magnitudes fainter than \host, making its flux contribution negligible given Kepler's photometric precision.

We show PDCSAP data containing the four transits of \plb\ and several transits of \plc\ in Figure~\ref{fig:nondetrended_transits}. The only additional detrending applied to the PDCSAP flux at this stage was a simple outlier removal. The light curves showed low frequency signals, likely due to stellar variability, but these were handled later in the modeling (Section~\ref{sec:model}). In total, Kepler observed four transits of \plb\ and over 50 transits of \plc. 

The \host\ system was included in the transit timing variation (TTV) analysis of \citet{Holczer2016}. The average significance (i.e., measurement divided by uncertainty) of the TTVs in the transits of both planets was less than unity. As a result, we did not account for possible TTVs when modeling the ephemeris of the \host\ planets.

For completeness, it is worth mentioning that the Transiting Exoplanet Survey Satellite (TESS) mission has also observed \host. However, at the time of writing, it has not observed any transits of \plb. We choose to not include TESS data for \plc\ owing to its lower photometric sensitivity relative to Kepler.


\subsection{Spectroscopic data from HIRES}\label{sec:hires}

We acquired 20 spectra of \host\ from the W. M. Keck Observatory using the High Resolution Echelle Spectrometer \citep[HIRES;][]{Vogt1994} over a span of 8.5~years between June 2013 and November 2021. Precise radial velocities (RVs) for hundreds of stars using the iodine cell with HIRES have been shown to have no systematic errors over 20 year timescales \citep{Rosenthal2021}. 19 of these 20 spectra were acquired with a heated I$_2$ cell in the light path, which imprinted a set of reference spectral lines. These I$_2$ lines enable wavelength calibration and tracking of the instrument profile for each observation. In the forward model, the line spread function is modeled as a series of Gaussians. The full model includes the line spread function parameters, the star’s radial velocity, wavelength solution, dispersion and a few other non-critical variables. The spectrum is divided into 700 individual 2-$\AA$ sub-sections of spectrum from roughly $5000-6200 \, \AA$ and the highest weight is assigned to the best performing sub-sections. There are no identified chromatic or other systematic errors that cause linear trends.

The HIRES I$_2$-in spectra were collected with the C2 decker, which spans $14.0" \times 0.87"$ in the spatial and dispersion direction, respectively. The spectra have a resolution of 60,000 at 5500 $\AA$. The minimum, median, and maximum signal-to-noise ratios (SNR) per pixel at 5500 $\AA$ are 51, 72 and 105. The other observation was taken without the I$_2$ cell (decker dimensions $14.0" \times 0.5"$) and had a higher S/N (126). This spectrum was used as a template in the forward modeling procedure that yielded the precise RV corresponding to each I$_2$-in observation \citep{Butler1996, Howard2010, Fulton2015b}. We limit the spectral extraction width to 14 pixels ($2.7"$) from the center of the star and as a result the $\sim 8$ magnitude fainter nearby star is not in the extracted spectrum (see \S~\ref{sec:companion}). The precise RVs for \host\ are listed in Table~\ref{tab:rvs}. The RV errors in Table~\ref{tab:rvs} are internal errors from the I2-RV pipeline.

The wavelength coverage of HIRES ($\sim$360--900~nm) enables the measurement of the $S_{\rm HK}$ stellar activity indicator from the Ca II H and K lines \citep{Wright2004,Isaacson2010}. We include this indicator alongside the RVs in Table~\ref{tab:rvs}.

\begin{deluxetable}{ccc}
\tablecaption{RV Measurements of \host\ From Keck-HIRES. \label{tab:rvs}}
\tablehead{
  \colhead{BJD$_{\rm TDB}$} & 
  \colhead{RV (m s$^{-1}$)} &
  \colhead{$S_{\rm HK}$}}
\startdata
$2456449.90146$ & $37.8\pm3.2$ & $0.137\pm0.001$ \\
$2456484.90839$ & $36.3\pm2.8$ & $0.129\pm0.001$ \\
$2456513.84738$ & $45.3\pm3.0$ & $0.128\pm0.001$ \\
$2456524.76180$ & $21.7\pm2.8$ & $0.128\pm0.001$ \\
$2456532.79213$ & $28.8\pm2.6$ & $0.129\pm0.001$ \\
$2457200.97212$ & $33.1\pm3.6$ & $0.130\pm0.001$ \\
$2458294.92664$ & $11.8\pm3.4$ & $0.137\pm0.001$ \\
$2458329.90056$ & $-10.1\pm3.6$ & $0.132\pm0.001$ \\
$2458389.76351$ & $2.9\pm3.7$  & $0.133\pm0.001$ \\
$2458632.87369$ & $-18.1\pm4.2$ & $0.130\pm0.001$ \\
$2458714.89043$ & $10.0\pm3.3$ & $0.128\pm0.001$ \\
$2458787.74592$ & $-11.7\pm3.8$ & $0.122\pm0.001$ \\
$2459038.93765$ & $8.9\pm3.7$ & $0.130\pm0.001$ \\
$2459099.87107$ & $-20.2\pm4.2$ & $0.118\pm0.001$ \\
$2459296.13499$ & $-19.2\pm4.2$ & $0.149\pm0.001$ \\
$2459421.02970$ & $-33.1\pm4.7$ & $0.079\pm0.001$ \\
$2459445.99461$ & $-41.3\pm4.1$ & $0.120\pm0.001$ \\
$2459470.84732$ & $-40.0\pm3.4$ & $0.130\pm0.001$ \\
$2459546.76795$ & $-45.2\pm5.5$ & $0.123\pm0.001$ \\
\enddata
\end{deluxetable}


\section{Modeling Kepler-511 System Parameters}\label{sec:model}

\subsection{Stellar parameters}
\label{sec:stellar_prop}

\begin{deluxetable}{llc}
\tabletypesize{\scriptsize}
\tablecaption{Median Values and 68\% Confidence Interval for \host\ Stellar Parameters}
\tablehead{\colhead{~~~Parameter} & \colhead{Units} & \multicolumn{1}{c}{Values}}
\startdata
~~~~$M_*$ &Mass (\msun) &$1.024^{+0.070}_{-0.060}$\\
~~~~$R_*$ &Radius (\rsun) &$1.785^{+0.063}_{-0.061}$\\
~~~~$L_*$ &Luminosity (\lsun) &$3.49\pm0.20$\\
~~~~$F_{Bol}$ &Bolometric Flux (cgs) &$2.66\times10^{-10}\;^{+1.5\times10^{-11}}_{-1.4\times10^{-11}}$\\
~~~~$\rho_*$ &Density (cgs) &$0.253^{+0.035}_{-0.029}$\\
~~~~$\log{g}$ &Surface gravity (cgs) &$3.945^{+0.044}_{-0.040}$\\
~~~~$T_{\rm eff}$ &Effective Temperature (K) &$5902^{+88}_{-86}$\\
~~~~$[{\rm Fe/H}]$ &Metallicity (dex) &$-0.356^{+0.064}_{-0.060}$\\
~~~~$[{\rm Fe/H}]_{0}$ &Initial Metallicity$^{1}$  &$-0.264^{+0.082}_{-0.079}$\\
~~~~$Age$ &Age (Gyr) &$7.5^{+1.7}_{-1.5}$\\
~~~~$EEP$ &Equal Evolutionary Phase$^{2}$  &$456.7^{+3.3}_{-4.9}$\\
~~~~$A_V$ &V-band extinction (mag) &$0.103^{+0.062}_{-0.066}$\\
~~~~$\sigma_{SED}$ &SED error scaling  &$1.49^{+0.59}_{-0.36}$\\
~~~~$\varpi$ &Parallax (mas)$^{3}$ &$1.544\pm0.016$\\
~~~~$d$ &Distance (pc) &$647.8^{+6.8}_{-6.6}$\\
\enddata
\label{tab:stellar}
\tablenotetext{}{See Table 3 in \citet{Eastman2019} for a description of all parameters.}
\tablenotetext{}{Catalogue ID for Kepler-511: KOI-289, KIC-10386922, 2MASS J18514696+4734295, Gaia DR3 2107644188496163200}
\tablenotetext{1}{The metallicity of the star at birth.}
\tablenotetext{2}{Corresponds to static points in a star's evolutionary history. See \citet{Dotter2016}.}
\tablenotetext{3}{The fitted value is in good agreement with the Gaia DR3 parallax of $1.529 \pm 0.010$.}
\end{deluxetable}

\begin{figure}
    \centering
    \includegraphics[width=\linewidth]{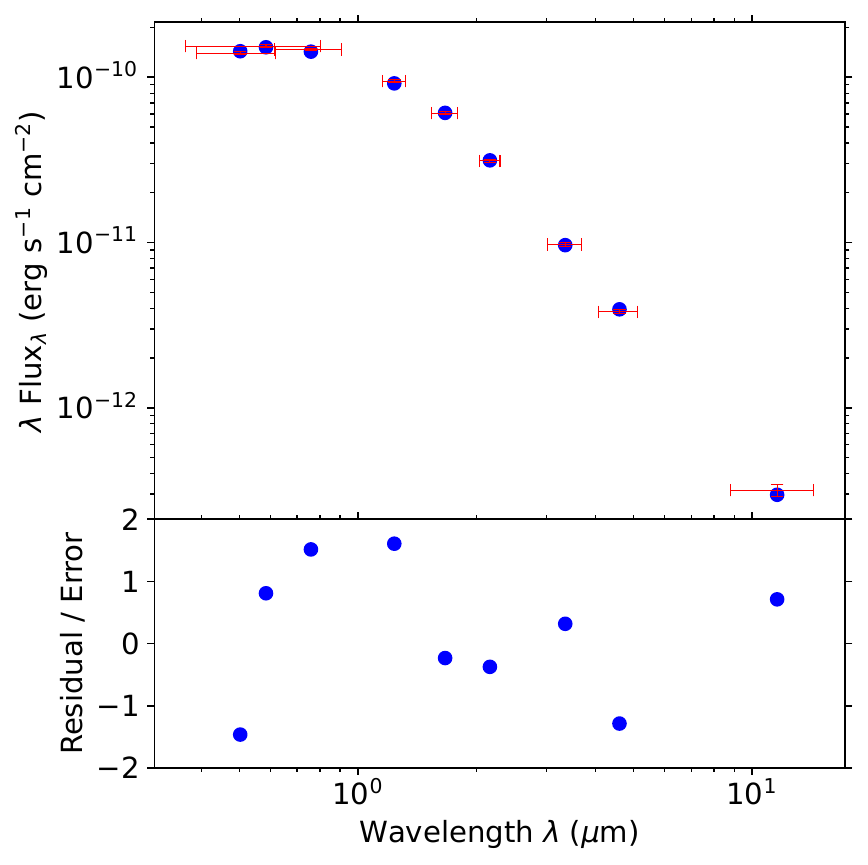}
    \caption{Top: The spectral energy distribution for \host\ from 2MASS, Gaia, and WISE in red along with the maximum likelihood model from our \textsf{EXOFASTv2} fit in blue. Bottom: Residuals of the fit normalized by the uncertainties in the SED measurements.}
    \label{fig:sed}
\end{figure}

We processed the high S/N template spectrum (Section~\ref{sec:hires}) of \host\ with \textsf{SpecMatch} \citep{Petigura2017b} to determine its spectroscopic properties: stellar effective temperature $T_{\rm eff} = 5855\pm100$~K, iron abundance [Fe/H] = $-0.36\pm0.06$~dex, surface gravity $\log g = 4.10\pm0.10$, and rotational velocity $v\sin{i} = 1.8\pm1.0$~km~s$^{-1}$. We then treated these values of $T_{\rm eff}$ and [Fe/H] as normal priors in an \textsf{EXOFASTv2} fit between stellar evolution models and archival broadband photometry \citep{Eastman2013,Eastman2019}. The \textsf{SpecMatch} results were not used to place a prior on $\log g$ as it is only weakly constrained by spectroscopic data and placing such a prior on it can bias fitted stellar properties \citep{Torres2012}. We point the reader to \citet{Eastman2019} for a complete description of the \textsf{EXOFASTv2} stellar evolution modeling. Briefly, \textsf{EXOFASTv2} interpolates a grid of MIST isochrones \citep{Paxton2011,Paxton2013,Paxton2015,Dotter2016,Choi2016} and precalculated bolometric corrections to derive basic stellar parameters and model the host star's spectral energy distribution (SED). Our \textsf{EXOFASTv2} fit included a uniform prior on reddening ($A_V \in [0, 0.19778]$) from galactic dust maps \citep{Schlafly2011} and parallax ($\varpi = 1.544 \pm 0.198$~mas) from Gaia Data Release (DR)~2 \citep{Gaia2016, Gaia2018}. We also imposed noise floors on $T_{\rm eff}$ and bolometric flux consistent with the findings of \citet{Tayar2020} to account for systematic uncertainties in the MIST models and to avoid overconstraining our stellar---and thereby planetary---parameters. The \textsf{EXOFASTv2} fit proceeded until convergence, which was assessed for each fitted parameter using the Gelman-Rubin statistic ($<$1.01) and the number of independent draws from the posterior ($>$1000). The SED of \host\ is shown in Figure~\ref{fig:sed}. As is default behavior for \textsf{EXOFASTv2}, the atmosphere model is not shown because this atmosphere was not used directly. Nonetheless, the residuals suggest that the fit is reasonable in each bandpass. The final stellar parameters are listed in Table~\ref{tab:stellar}. The fitted log~$g$ is compatible with the spectroscopic value at the 1.4~$\sigma$ level. We also verify that the fitted set of stellar parameters reproduces the star's observed luminosity using a different grid of stellar models (\textsf{PARSEC}, \citealt{Bressan2012}).

\host's luminosity and inferred stellar parameters imply that it is evolving off the main sequence. The broadband photometry of \host\ indicates that the star is significantly more luminous than the Sun\footnote{The exact luminosity does depend on the photometric points chosen for the fit. In their large scale studies of planet-hosting stars, \cite{Berger2020} and \cite{Berger2023} chose 2MASS + SDSS $g$-band photometry and the Gaia $G_{\rm BP}$ and $G_{\rm RP}$ photometry for \host\ and find that it is $2.75^{+0.20}_{-0.19}$ and $3.01^{+0.13}_{-0.12}$ times as luminous as the Sun, respectively. We use a wider range of photometric measurements for \host, which should provide a more reliable estimate of its stellar properties}. One potential confounding possibility for the source's high luminosity is the presence of an unresolved companion. \cite{Kraus2016} reported a potential companion with a $\Delta K$-band magnitude of 0.189 and a separation of $0\farcs017$ based on observations from NIRC2 on the Keck II telescope. Such a small $K$-band contrast requires the companion to be nearly equal mass and the small separation corresponds to a mere $\sim 11$ au in projected separation. The presence of such a binary companion would complicate our ability to place constraints on the properties of the two stars and consequently any planets around them (although the effect on planetary radii would be much smaller than the typical dilution correction, see end of this section).

To resolve this issue, we closely examined various lines of evidence that support or reject the hypothesis of a nearby companion of \host.
\begin{enumerate}
    \item We note that the reported detection pushed the resolving power of the NIRC2 instrument and that the companion has not been confidently recovered in follow up imaging observations of the system (\citealt{Dupuy2022}, A. Kraus, private communication). This could be due to orbital motion.
    \item However, the RV measurements of the star show a long-term trend of only $\sim 77$ m/s over the course of 8.5 years. The presumptive nearly-equal mass stellar companion would have to be on a highly improbable nearly face-on orbit to produce this relatively small RV trend. Such an orbit would also suggest that the companion should be detectable at other epochs, in contrast to follow up observations.
    \item The high-resolution template spectrum of the star obtained as part of the RV observations does not contain any hint of the presence of two stars and is well fitted by a single star template (Appendix~\ref{sec:kolbl_analysis}, \citealt{Kolbl2015}). Stars with an RV separation of $\geq \pm 10$ km/s and flux level $> 1\%$ of the primary are ruled out. If the binary companion is spectroscopically unresolved, we would expect the line profiles of the source to vary with time due to stellar orbital motion. However, the $\chi^2$ and residuals of RV spectra fits do not exhibit any meaningful variation with epoch. 
    \item The Gaia RUWE value is merely 1.03 for Kepler-511, which is below the threshold of 1.4 that typically indicates the presence of an unresolved companion. The excess astrometric noise for Kepler-511 is 0.06 mas ($4.8 \sigma$ level). This is much smaller than what we would expect from a face-on nearly equal mass companion. Using the derived mass ($q = 0.983$) and luminosity ratios (adjusted to optical, $l \sim 1/1.3$) from \cite{Kraus2016}, we would expect the center-of-light to orbit around the center-of-mass \citep{Penoyre2020} with a separation of $|q - l| / (1 + q) / (1 + l) \times$ 17 mas $\sim$ 1.04 mas $>>$ 0.06 mas. The astrometric noise is likely due to the presence of planets and a distant sub-stellar companion around the star (\S~\ref{sec:companion}). We also note that the astrometric excess noise may be higher in the early data releases from Gaia because it absorbs many different noise sources and that RUWE is a more reliable measure of the goodness of fit compared to the astrometric excess noise because RUWE is corrected for calibration error but astrometric excess noise is not \citep{Lindegren2021}.
    \item The kinematic properties of the Kepler-511 system suggest that it is a member of the thick disk population with a thick disk to thin disk membership probability ratio of $\sim 800$ \citep{Chen2021}. This is compatible with the stellar age implied by ischrone fitting.
\end{enumerate}

All of these considerations lead us to conclude that \host\ is likely a slightly-evolved single star rather than a nearly equal mass binary. If the system is a binary, both stars would still need to be slightly evolved since their combined luminosity = $3.49 \pm 0.20 \, L_{\odot}$ and their luminosity ratio in $K$-band is 1.19. We can estimate the magnitude of a secondary star's effect on the planets' radii. In our case, the correction would be significantly smaller than one would naively assume because accounting for a binary companion changes both the radius and the flux of the planet-hosting star. If $\delta$ and $l = l_{\rm secondary} / l_{\rm primary} < 1$ are respectively the transit depth and the luminosity ratio of the two stars and assuming the planets orbit the primary star (easily extendable to the converse case), the corrected transit depth would be
\begin{equation}
    \delta_{\rm corrected} = \delta_{\rm observed} \, (1 + l).
\end{equation}
To obtain the planet radius, we would multiply $\sqrt{\delta}$ with the primary star's radius $R_{\rm primary}$:
\begin{align}
    R_{\rm p, corrected} & = \sqrt{\delta_{\rm observed} \, (1 + l)} \, R_{\rm primary} \nonumber \\
     & = \sqrt{\delta_{\rm observed} \, (1 + l)} \, \sqrt{\frac{L_{\rm total}}{4 \pi \sigma_{\rm SB} T_{\rm primary}^4 (1 + l)}} \nonumber \\
      & = \sqrt{\frac{\delta_{\rm observed} L_{\rm total}}{4 \pi \sigma_{\rm SB} T_{\rm primary}^4}},
\end{align}
where $L_{\rm total}$ is the total luminosity of the system, $T_{\rm primary}$ is the effective temperature of the primary planet hosting star, and $\sigma_{\rm SB}$ is the Stefan-Boltzmann constant. If $T_{\rm single}$ denotes the effective temperature of a single evolved star that accounts for the total observed luminosity, the ratio of the inferred planetary radii in the binary scenario vs the single star scenario would be $ = (T_{\rm single} / T_{\rm primary})^2$. If the nearly equal mass companion is real and unresolved in photometric data, we expect $T_{\rm primary} \sim T_{\rm single}$ and therefore the effect of binarity on the planets' radii should be much smaller than the typical dilution correction of $\sqrt{1 + l}$. Continued RV monitoring and future direct imaging observations are essential for illuminating  the nature of this system but until the presence of a luminous companion is vindicated, we proceed assuming that \host\ is a single evolved star.

\begin{figure}
    \centering
    \includegraphics[width=\columnwidth]{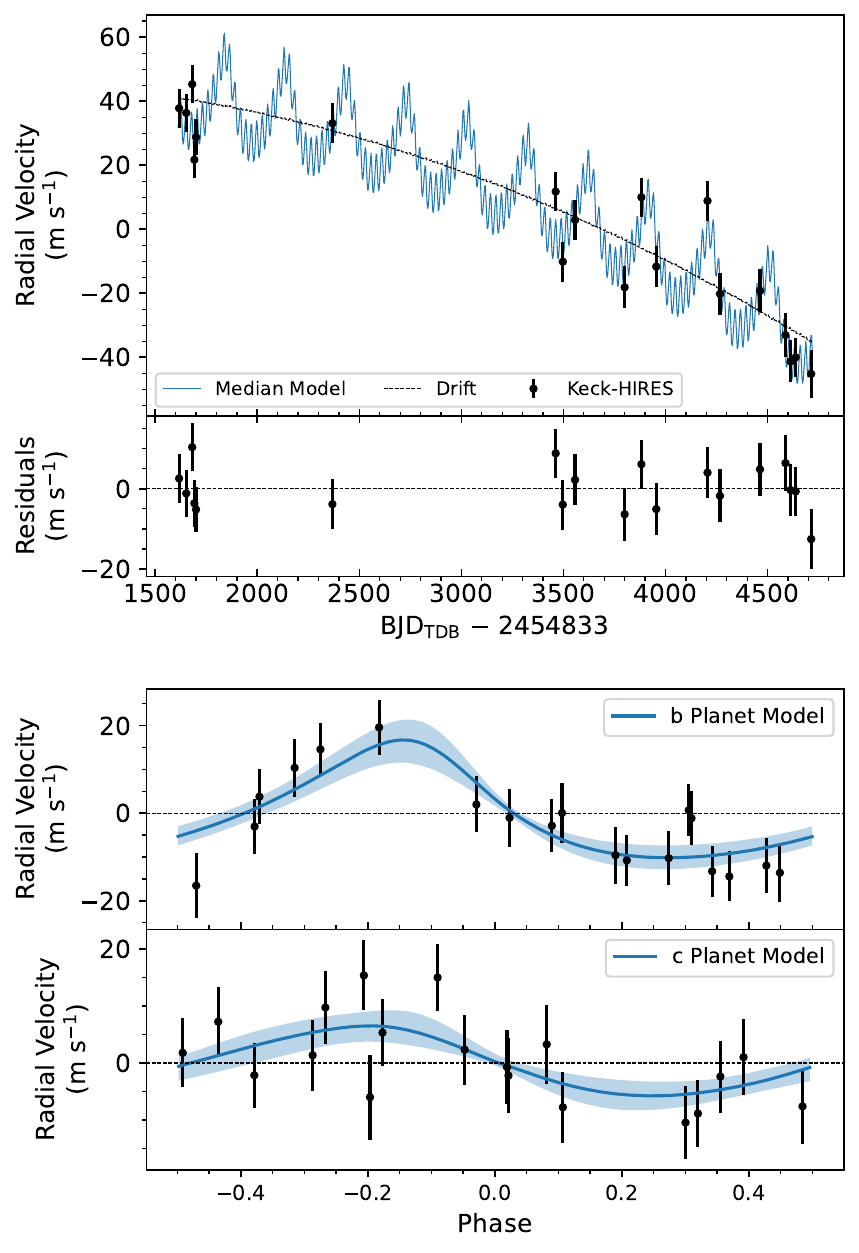}
    \caption{Top: time-series RV data from Keck-HIRES (black points) and median RV model (blue line). The need to include a long-term trend (dashed black line) in addition to the two planetary Keplerlian signals is evident. The $\chi^2$ of the RV model to the data (19 data points) with and without the RV jitter term is 15 and 45, respectively. Bottom: individual components of the RV for each planet folded on the best fit ephemeris such that a phase of zero is the conjuntion time (Table~\ref{tab:stellar}). The blue lines are the median models and the blue shaded regions are the 68\% confidence intervals.}
    \label{fig:rv}
\end{figure}

\subsection{Planetary parameters}

We used \textsf{exoplanet} \citep{ForemanMackey2021} to jointly model the \host\ transit and RV data. We parameterized the transit models with the orbital period, the ratio between planetary and stellar radius, the stellar density, and quadratic limb darkening parameters following \citet{Kipping2013b}. We used our fit to the broadband photometry of the star to place a normal prior on the stellar density (Table~\ref{tab:stellar}) in our transit fit. A fit without a prior on the stellar density produced a higher likelihood ($\Delta$BIC = 14) and higher stellar density that agrees with the SED fitted value at the 2$~\sigma$ level. However, this difference in likelihood is driven entirely by the small number of RV data points, which biases the stellar density to higher values (see Appendix~\ref{sec:stellar_density}). The normal prior on stellar density is therefore placed to alleviate this issue, in line with previous studies \citep{Espinoza2019}. The transit model also included a Gaussian Process (GP) kernel corresponding to a damped simple harmonic oscillator meant to model the stellar variability signal in the photometry \citep[e.g.,][]{ForemanMackey2021}. The short and long cadence data were treated with separate GP models, and only a subset of the Kepler data during and immediately surrounding the transits was fitted. For the RV model, we included terms for a quadratic trend, which were necessary based upon inspection of the time-series data (Figure~\ref{fig:rv}, \S~\ref{sec:companion}).

\begin{figure*}
    \centering
    \includegraphics[width=\textwidth]{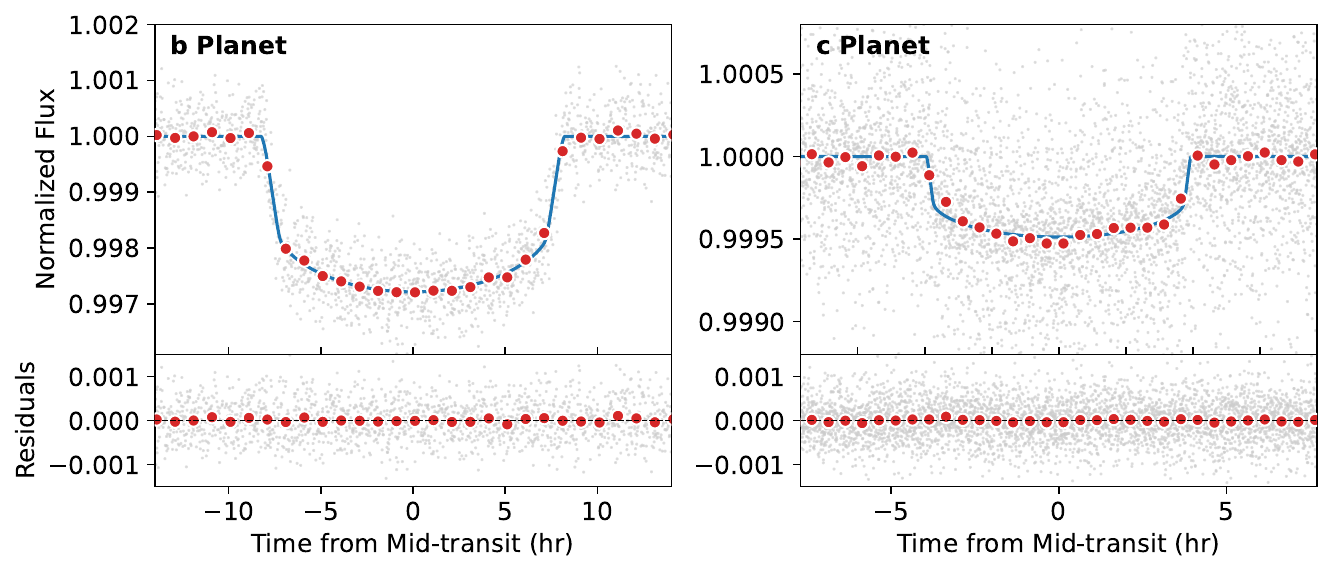}
    \caption{Kepler data and the median transit model, folded on the best fit ephemeris (Table~\ref{tab:planet}), for both \host\ planets. Gray points represent individual frames, and red points are 1~hour and 0.5~hour bins for \plb\ and \plc, respectively. The RMS error on the residuals is 414 ppm.}
    \label{fig:transits}
\end{figure*}

The \textsf{exoplanet} fit proceeded in two parts. First, there was an optimization of the fitted parameters to identify the local maximum a posteriori (MAP) point of the model. This MAP model included the GP for the photometry as well as the transits. We ran the MAP model several times, adjusting the starting points of the parameters and the order in which the parameters were optimized. Each time, we visually inspected the residuals with the transit and RV data until we were satisfied that the MAP solution was representative of the transit and RV data. We derived the photometric variation caused by stellar variability from this optimization and subtracted it from the transit light curves to effectively flatten the photometry. The stellar rotation period inferred via this maximum likelihood method was 25.1~d.

Second, we launched the Hamiltonian Monte Carlo (HMC) routine to generate posterior predictive samples from the joint transit and RV model using the flattened photometry. The parameter estimation proceeded with 1,500 tuning steps before making 6,000 draws of the posterior. At that point, each fitted parameter had a Gelman-Rubin statistic of $<$1.01 and an effective sample size of $\gtrsim$1,000, which we took to demonstrate convergence. 

The median transit models, folded on the ephemeris of each planet, are shown in Figure~\ref{fig:transits}. The median and 68\% confidence interval of the RV model are shown in Figure~\ref{fig:rv}. The fitted parameters along with a few informative derived parameters are listed in Table~\ref{tab:planet}. In deriving physical parameters from relative ones, we used the stellar properties and uncertainties listed in Table~\ref{tab:stellar}. \host\ b and c have measured masses of $0.44^{+0.11}_{-0.12}$ (4$\sigma$) and $0.100^{+0.036}_{-0.039}$ ($2.6 \sigma$) Jupiter masses respectively.

\input{planet_table_RHO_STAR}

\subsection{A distant companion in \host\ system?}\label{sec:companion}

\begin{figure}
    \centering
    \includegraphics[width=\linewidth]{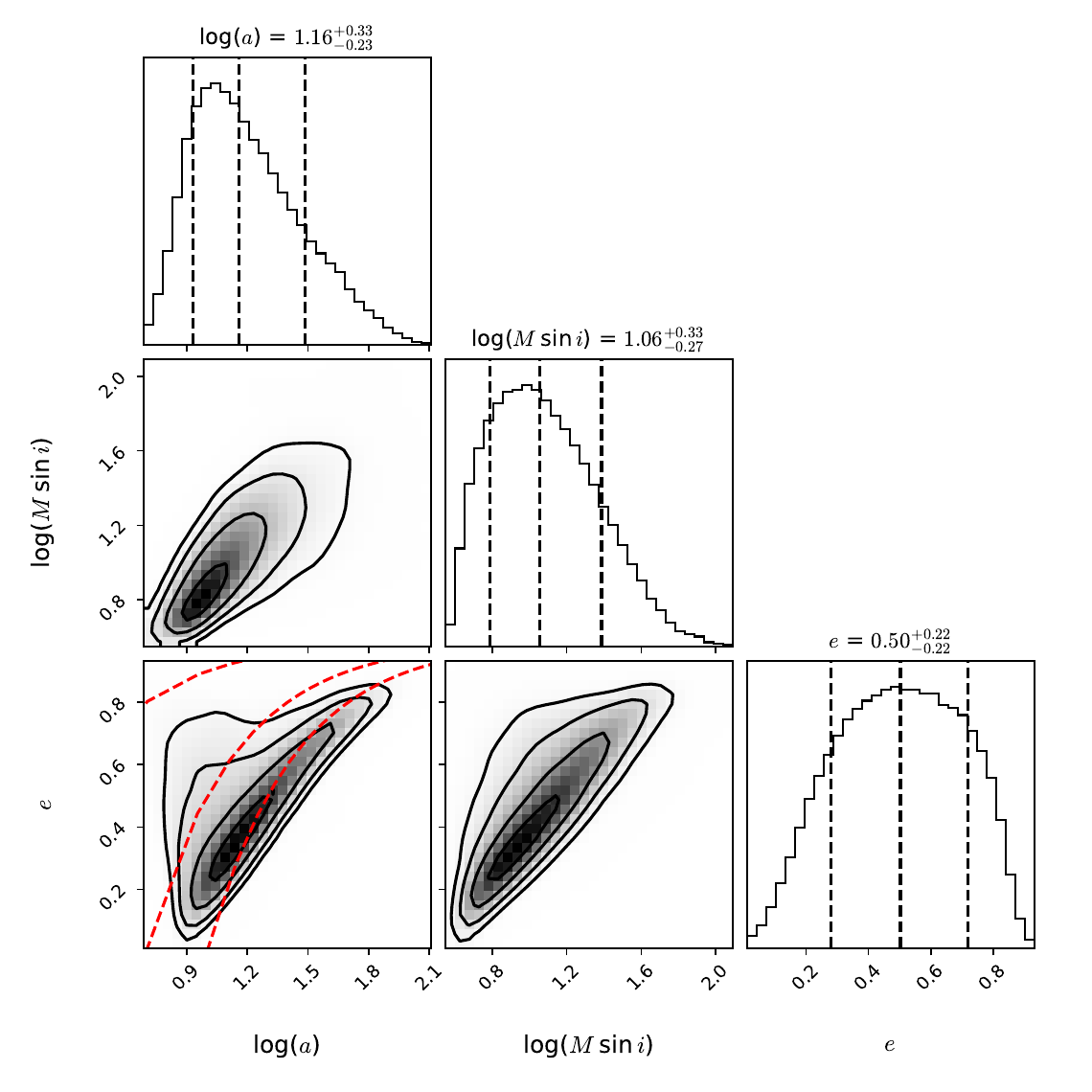}
    \includegraphics[width=\linewidth]{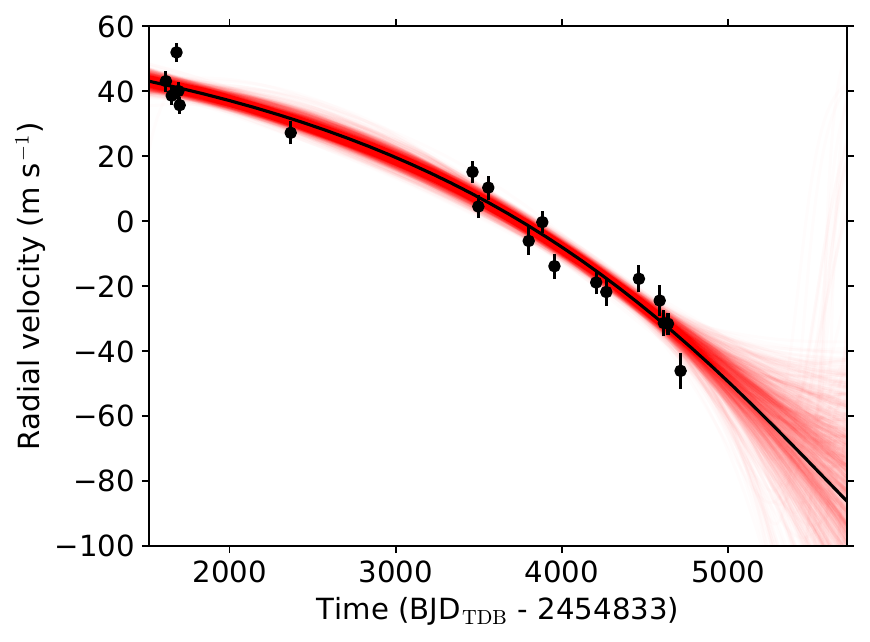}
    \caption{Top: Corner plot for the constraints on the semimajor axis (AU), $M~{\rm sin}i$ ($M_{\rm Jup}$), and eccentricity of the distant companion. The red dashed lines in the eccentricity - semimajor axis 2D histogram mark pericenter values of 1, 5, and 10 au.  Bottom: Black points mark the RV data with the best fit model for the two transiting planets subtracted out. The black curve shows the best fit model to the long term trend and the red curves correspond to 1000 models from randomly drawn parameters from the posterior.}
    \label{fig:trend_fitting}
\end{figure}

Possibly adding to the peculiarity of the \host\ system is the presence of a large RV drift ($\sim 77$ m/s) present over the 8.5~year observational baseline (Figure~\ref{fig:rv}). This RV trend is unlikely to be due to stellar activity as its amplitude is larger than what one might expect from the activity \citep[e.g.,][]{Lovis2011} and because there is little correlation (Pearson correlation coefficient of 0.35) between the RVs and the $S_{\rm HK}$ values measured from the Ca II H and K lines for the Keck-HIRES RVs (Table~\ref{tab:rvs}) \citep[e.g.,][]{Diaz2016, Butler2017, Rosenthal2021}. 

The most plausible  hypothesis for the cause of this trend is a distant companion orbiting far beyond \plb. Such a companion would have important implications for the formation of the entire system, as planet-star and planet-planet interaction alike can excite eccentricity of inner planets \citep[e.g.,][]{Wu2003,Lithwick2011b}. A direct imaging campaign with the NIRC2 instrument on the Keck II telescope that targeted \host for resolved stellar companions claimed two detections \citep{Kraus2016, Furlan2017a}: the first with a $\Delta K$-band magnitude of 0.189 and separation of $0\farcs017$, and the second with a $\Delta K$-band magnitude of 7.796 and separation of $3\farcs26$. We presented multiple lines of reasoning to argue that the closer imaged companion is a false positive in Section~\ref{sec:stellar_prop}. Such a small $K$-band contrast would imply that this purported companion is nearly equal mass as the primary star. We rule out companions brighter than 1\% of the primary's brightness with RV separations of $\geq 10$ km s$^{-1}$ (\S~\ref{sec:kolbl_analysis}). The observed RV trend would thus require the companion to be on a highly improbable face-on orbit. If it is on such an orbit, it should be detectable at multiple epochs but it has not been observed again since its initial detection \citep{Dupuy2022}. As for the companion at $3\farcs26$, recent Gaia astrometry identifies it as Gaia DR~3 2107644188495101440, which is clearly not bound to \host\ based on proper motion \citep{Gaia2022}. 

\begin{figure*}
    \centering
    \includegraphics[width=0.49\linewidth]{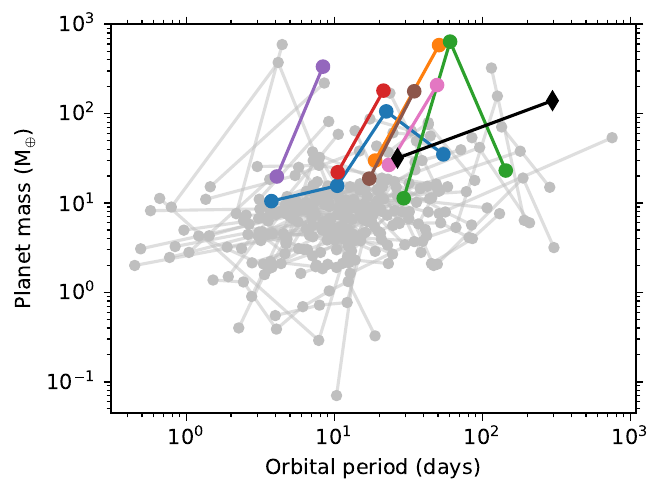}
    \includegraphics[width=0.49\linewidth]{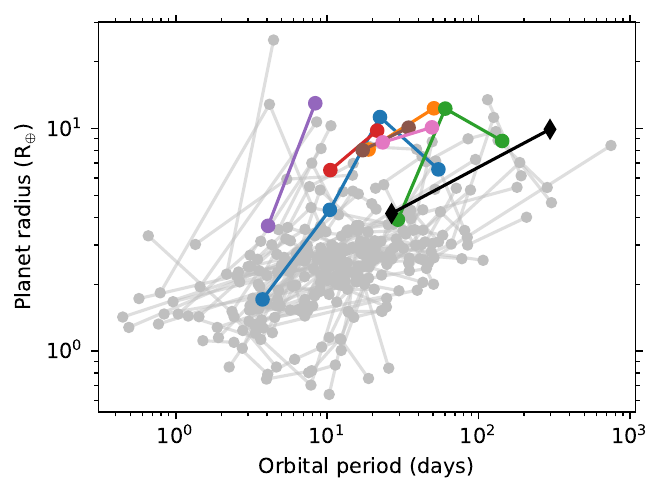}
    \includegraphics[width=0.49\linewidth]{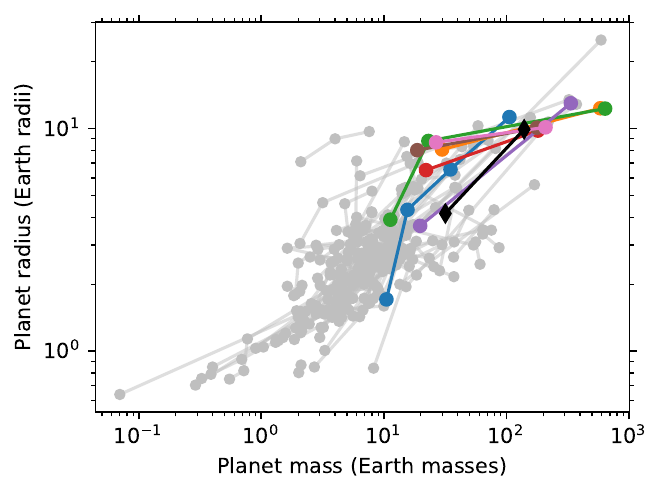}
    \caption{Multi-planet systems with planets that have measurements for both their masses and radii are shown (data from the exoplanet archive). Systems that do not have a single planet with mass $\geq 95 M_{\oplus}$ or any planet with $17 M_{\oplus} <$ mass $< 95 M_{\oplus}$ are grayed out. The \host\ planets are shown with black diamonds. \plb\ is the longest orbital period planet in such a sample. Gas rich giants ($> 95 M_{\oplus}$) are typically accompanied by planets that are significantly less massive than \plc. Some planets that do have masses similar to \plc\ are typically much larger than it, indicating that they are much more gas rich compared to \plc.}
    \label{fig:kepler-511-context}
\end{figure*}

The RV drift shows significant curvature (the quadratic trend coefficient is non-zero at $\sim 4 \sigma$ level), possibly suggesting that the observations sampled a portion of the distant companion's orbit near quadrature and allowing us to draw useful constraints on the distant companion's properties. We put constraints on the companion properties by subtracting the best fit solutions for planets b and c from the RV data and fitting the residual long-term trend with \textsf{RadVel}. This procedure does not allow us to account for the covariance between properties of the transiting planets and the distant companion. However, given the limited number of RV measurements, a three planet fit is unfeasible due to the large number of free parameters. Our constraints on the distant companion are likely to be tighter than the data merit but still useful for gauging the companion's properties and guiding follow up RV campaigns. The top panel of Figure~\ref{fig:trend_fitting} shows the posteriors for the distant companion's semi-major axis, minimum mass ($M \, {\rm sin} i$), and eccentricity. The RV trend, best fitting model, and 1000 sample models are shown in the bottom panel of Figure~\ref{fig:trend_fitting}.

The distant companion in the Kepler-511 system may be planetary or stellar in nature. Given that we sample its orbit only partially, our constraints on its semi-major axis, mass, and eccentricity are highly correlated. Combinations of eccentricity and semi-major axis that lead to pericenter values (1, 5, and 10 au marked with red dashed lines in top panel of Figure~\ref{fig:trend_fitting}) close to the inner planets' orbits are implausible as they would violate the long-term stability of the system. These constraints also confirm our earlier analysis regarding the plausibility of a nearly equal mass stellar companion around Kepler-511: it would need to be on a nearly face on orbit to match the observed RV trend. The distant companion is more likely to be a sub-stellar object or a very low mass star with a semi-major axis of $\sim 10 - 100$ au potentially on an eccentric orbit. Continued RV follow up would be extremely useful for characterizing the properties of this companion and for understanding the context in which the inner planets formed and evolved.


\section{The Kepler-511 planets}\label{sec:res}

Based upon the aforementioned photometric and spectroscopic data acquired of the \host\ system, we find that \plb\ and \plc\ both have masses well within the planetary regime, thereby dynamically confirming them as genuine exoplanets. Figure~\ref{fig:kepler-511-context} shows planets b and c in context with other planets in multiplanet systems that have measured masses and radii. \plb\ joins the small but growing group of transiting cool giant exoplanets on orbits longer than $\sim$100~days with precisely measured masses \citep[e.g.,][]{Dalba2021a}. \plc\ presents as a dense Neptune-sized planet that may have a considerable gas envelope, just based on its mass and radius. 

\begin{figure*}
  \begin{center}
    \begin{tabular}{cc}
      \includegraphics[width=\columnwidth]{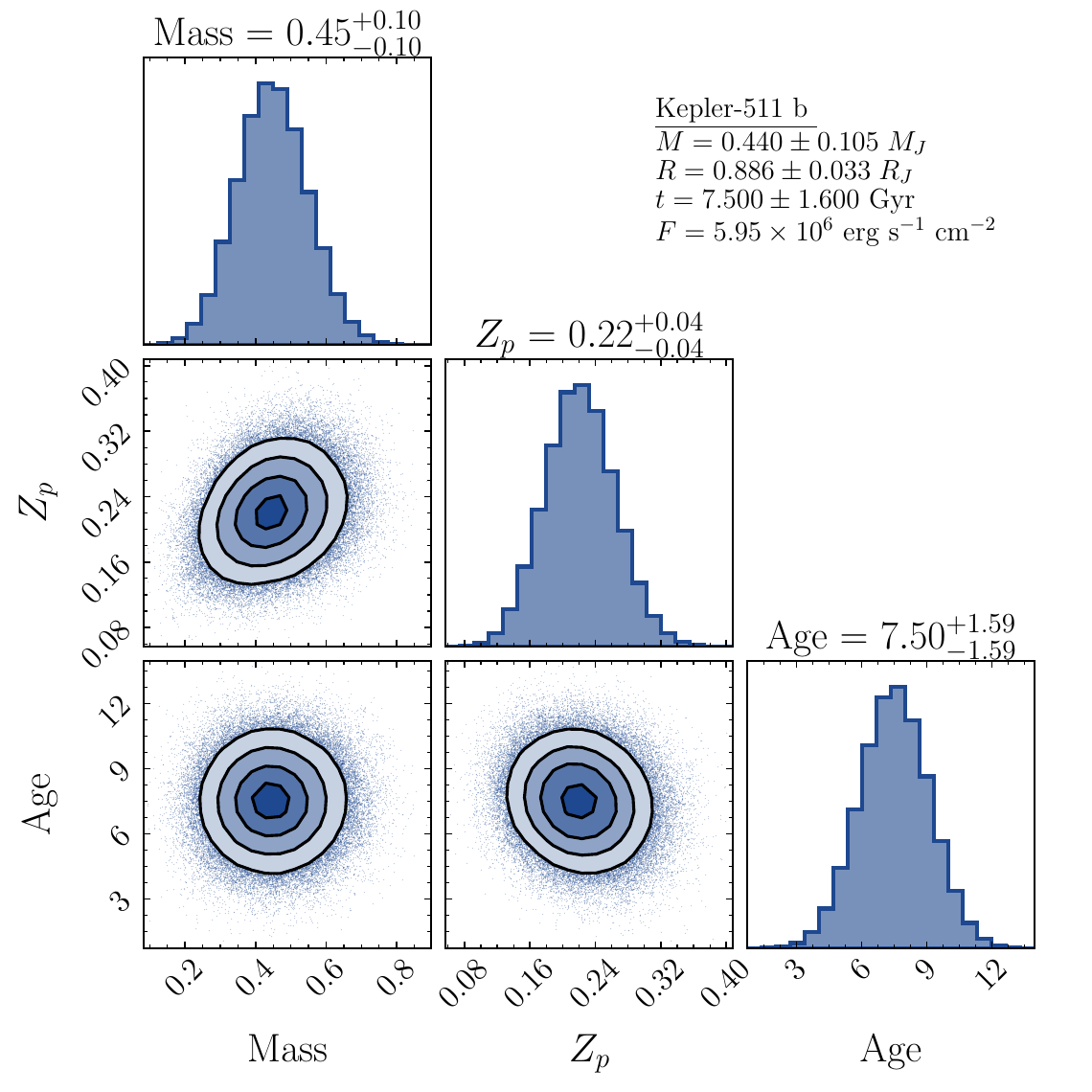} 
      \includegraphics[width=\columnwidth]{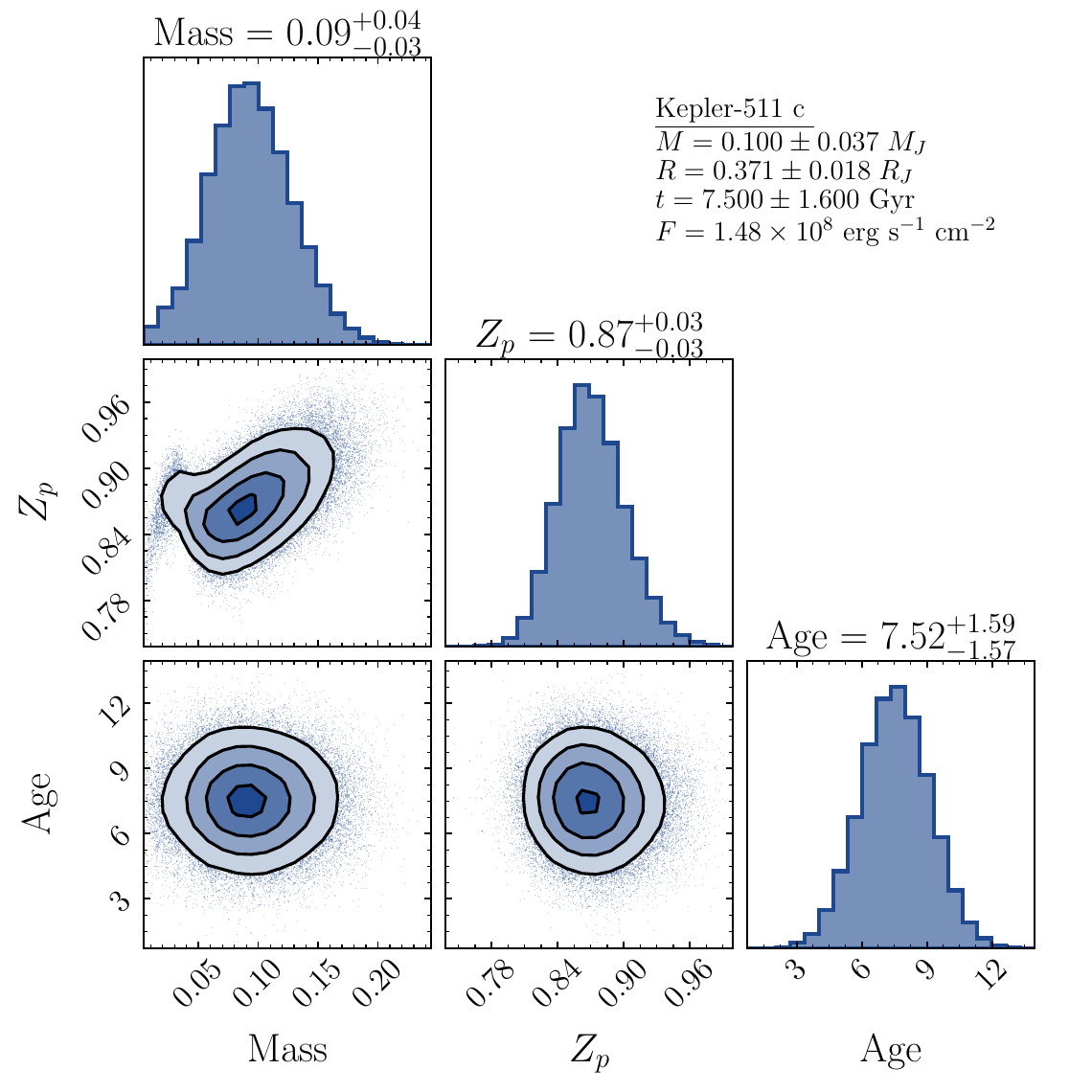} 
    \end{tabular}
  \end{center}
  \caption{Posterior probability distributions for bulk metallicity ($Z_p$) retrievals for \plb\ (left) and \plc\ (right). For Kepler-511 b, the inferred values of $Z_p$ do not show significant correlations with the measured planets masses or system age; for \plc, mass and $Z_p$ are positively correlated to fit the planet's observed radius (planet radius increases with mass at a constant metallicity in this planet mass range). By mass, \plb\ and \plc\ are 22\% and 87\% heavy elements.}
  \label{fig:metals}
\end{figure*}

Amongst multi-planet systems, the \host\ system is unique in that \plc\ is significantly more massive than the inner super-Earths that typically accompany cold giants \citep[e.g.,][]{Zhu2018, Bryan2019}. In addition, \plc\ is much smaller in size than planets of comparable mass in this multiplanet sample, indicating that it is not as gas rich as other similar mass planets that accompany outer giants. The only other system that appears qualitatively similar to the \host\ planets is the TOI-1130 system (\citealt{Borsato2024}, shown in purple in Figure~\ref{fig:kepler-511-context}), albeit at significantly shorter orbital periods). The combination of these properties makes the \host\ system interesting from a planet formation perspective. To explore the formation scenarios for the \host\ planets further, we quantified the bulk heavy element content of both planets.

\subsection{Planetary bulk metallicity}\label{sec:zp}

With a known mass, radius, and age for each planet in the \host\ system, we retrieved the total mass of heavy elements that are possibly present within these planets (i.e., their bulk metal mass $M_Z$). We use a 1D planetary evolution model coupled to a radiative atmosphere model in order to evolve the radius of a giant planet along its cooling curve at a given mass. At the age of the system, we adjust $M_Z$ to retrieve the known radius. The metals are assumed to be entirely in the core if $M_Z \leq 10 M_\oplus$. Any additional metal mass is incorporated into the envelope.  For the model atmospheres we use \citet{Fortney2007}, a non-grey two-stream model at solar metallicity, to determine the rate of heat flow out of the interior. The \host\ planets are cool enough to not be affected by radius inflation. Additional details of this method and the models and assumptions used are reported by \citet{Thorngren2016} and \citet{Thorngren2019a}. 

We found that the bulk metallicities ($Z_p \equiv M_Z/M_p$) of \plb\ and c are $0.22\pm0.04$ ($Z_p / Z_{\star} = 35\pm8$) and $0.87\pm0.03$ ($Z_p / Z_{\star} = 143\pm20$), respectively. Figure~\ref{fig:metals} shows the posteriors distributions for these metal retrievals and demonstrates that the fits are converged. Most parameters lack strong correlations with the exception of planet mass and metallicity for \plc. In \plc's mass range, planet radius increases with mass for constant metallicity and as a result, \plc's mass and metallicity are positively correlated to match the planet's observed radius. Both planets are above the average of, although still consistent with, the correlation between relative bulk metallicity and giant planet mass identified by \citet{Thorngren2016}.

Note that the uncertainties listed for the planet metallicity reflect only the statistical uncertainty arising from observational uncertainties in mass, radius and age -- they do not include theoretical uncertainty.  The latter is difficult to quantify rigorously but the dominant sources are the structuring of metals within a planet (e.g. a core-dominated vs a well-mixed interior), uncertainties in the equations of state, and to a lesser extent our assumptions about how the planet has evolved thermally.  For the first of these, moving metal from the envelope to the core in the model increases the inferred metallicity to some extent \citep[see][for further discussion of this modeling uncertainty]{Thorngren2016}.  For equation of state uncertainties, those of H/He are the most important and in particular the calculation of their adiabats \citep{Miguel2016}.

The measured bulk metallicities imply remarkably similar amounts ($31^{+12}_{-10}$~$M_{\oplus}$ and $26^{+11}_{-10}$~$M_{\oplus}$ respectively) of heavy elements in planets b and c despite the fact that they differ in mass by a factor of $\sim 4.5$. This difference suggests that \plb\ experienced runaway gas accretion while \plc\ did not. Such a scenario is peculiar given that ability of a planet to accrete gas is strongly dependent on the amount of heavy elements it contains. Despite having similar amounts of heavy elements, these planets have very different gas contents. This makes the Kepler-511 planets invaluable for studying giant planet formation history and gauging the role of other properties that control gas accretion.

It is interesting that two metal enriched giant planets formed around such a metal-poor host star ([Fe/H] = $-0.36$) given the observed correlation between stellar metallicity and giant planet occurrence \citep[e.g.,][]{Fischer2005}. The combined metal mass in the two planets stretches the initial disk mass one needs for their formation around such a low metallicity star. Assuming i) a 100\% efficiency of converting dust to these planetary cores, ii) the disk has the same metallicity as the host star, and iii) a dust to gas mass ratio of 0.01 for a solar metallicity star, 57 $M_{\oplus}$ of dust mass would require a gas disk that is at least 3.8\% the mass of the host star ($1.024 \, M_{\odot} \times 0.01 \times 10^{-0.356} \times 0.038 \sim 57 \, M_{\oplus}$). Given that efficiency of converting dust to planets is typically much smaller than 100\% \citep[e.g.,][]{Drazkowska2016, Chachan2023} and this estimate does not include the mass contained within the distant companion, the protostellar disk around Kepler-511 must have been $\gtrsim 10-20\%$ of the mass of the host star. Such a massive disk was likely prone to gravitational instability \citep{Boss1997}, which may have given birth to the star's distant sub-stellar companion. In the following section, we will consider how various theories of giant planet formation can explain the inferred bulk metallicities for the \host\ planets.


\section{Formation and Evolution Histories of the Kepler-511 planets}\label{sec:disc}

The planets in the \host\ system demonstrate the benefit of measuring precise masses and making bulk composition assessments for cool, giant exoplanets. Having only validated these planets (i.e., only measured their radii), our interpretation was limited to the fact that \host\ is a system of multiple giant planets. By measuring masses and inferring their bulk metallicities, we can now conclude that these two sibling planets underwent different accretion processes during formation. In the following sections, we consider three broad explanations for how \plb\ and \plc\ formed and why they are so different from each other. 

\subsection{Envelope mass loss history}

As the first explanation for the large composition difference between \plb\ and c, we consider whether XUV-driven mass loss could have removed a substantial portion of the inner planet's envelope. Considering the XUV mass loss models of \citet[][their Figure~5]{Lopez2012}, a planet with a bulk density of 2.7~g~cm$^{-3}$ and receiving an irradiation of $\sim$110~$S_{\oplus}$ should experience negligible mass loss on the order of 0.1~$M_{\oplus}$ per Gyr (0.3\% of \plc's current total mass per Gyr). To verify and refine this approximate result, we applied the models of \citet{Thorngren2023} to \plc. These models simulate the thermal and mass-loss evolution of giant planets using the stellar XUV evolution tracks of \citet{Johnstone2021} and the XUV-driven mass-loss models of \citet{Caldiroli2022}. 

For the planet's structure, we adopt a core mass of 10~$M_{\oplus}$ and adjust the envelope metallicity to 0.81 to match the observed radius (for a total $Z=0.87$). Evolving forward from 0.101~$M_{\rm J}$, we estimate that the planet has lost just 0.194~$M_{\oplus}$ of envelope material (0.67\% of \plc's current total mass) over its $\sim$7.5~Gyr. lifetime. We find an insignificant present-day mass loss rate of approximately 0.00906~$M_{\oplus}$~Gyr$^{-1}$ (0.03\% of \plc's current total mass). The estimates of mass loss are insensitive to the adopted core mass since the envelope metallicity will have to be adjusted accordingly to match the radius. For example, using a 25 M$_{\oplus}$ core results in the loss of 0.188~$M_{\oplus}$ over the planet's life. While the rate of mass loss may vary somewhat based on how active the star was in its early life (most of the mass-loss occurs in the first couple Gyr.), there is not a plausible value that would lead to substantial mass loss -- the planet is simply too massive and distant from its parent star.

\subsection{Differing envelope accretion histories}

The accretion of primordial envelopes is controlled by the mass of the planetary core, the metal content of the accreted gas, and to a lesser extent, the ambient nebular conditions \citep{Stevenson1982, Rafikov2006, Lee2014, Piso2015}. Given the remarkable similarities in the total metal content of the two planets, it is reasonable to first assume that the two planets had cores of similar masses. If the two cores are assumed to be 26~$M_{\earth}$ each, can differences in accreted material or nebular conditions account for their differing final masses and bulk densities? For this scenario to be plausible, \plb\ must attain a gas-to-core mass ratio (GCR) of $>0.5$ and subsequently undergo runaway accretion while \plc\ only reaches a GCR of $\sim1/6.7$ ($= (1 - Z_{\rm p}) / Z_{\rm p}$, assuming all the metals are in the core) before the dissipation of the protoplanetary disk. For such a scenario, we assume that the cores accreted for the same duration, which provides us with a useful limiting constraint on their accretion environment.

This difference is unlikely to be driven by differences in ambient gas density at the location of the two planets. The gas density at \plc's location would need to be a factor of $\sim 10^4$ lower compared to the density at \plb's location (GCR $\propto \Sigma^{0.12}$, \citealt{Lee2021}), which would require \plc's 26 $M_\oplus$ core to assemble only as the gas begins dissipating, which is similar to the merger scenario considered in \S~\ref{sec:mergers}. Gap opening is also unlikely to be effective at creating such large gas density contrasts (assuming $M_{\rm p} = 26 \, M_{\oplus}$, the gap depths would only differ by an order of magnitude for a factor of two difference in the disk aspect ratio, \citealt{Duffell2013, Fung2014}). We therefore focus on the formation of the \host\ planets assuming accretion of material with different dust content. The accretion of gas is mediated by the cooling of the planet's envelope \citep{Lee2015}, which in turn depends on the opacity at the innermost radiative-convective boundary (RCB) of the envelope. The innermost RCB of envelopes that have even a small amount of dust ($\gtrsim 0.01$ solar metallicity for ISM-like size distribution) is set by the hydrogen dissociation front, inside which H$^-$ opacity dominates. Metallic species accreted by a planet are the primary source of free electrons that create H$^-$ ions, and a large difference in the local disk metallicity (as can be probed with dust content) could therefore produce the divergent accretion histories of the \host\ planets \citep{Chachan2021}.

\begin{figure}
    \centering
    \includegraphics[width=\columnwidth]{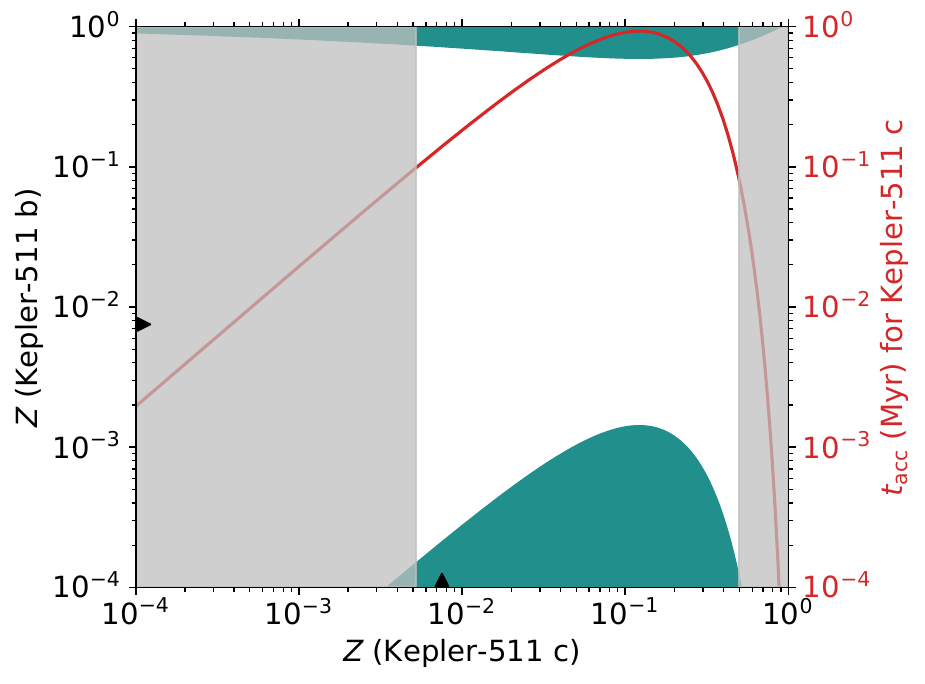}
    \caption{Bulk metallicity parameter space for \plb\ (y-axis) and \plc\ (x-axis). On both axes, the stellar metallicity is shown with the black triangles. The solid red line (corresponding to the right-side y-axis) shows the time ($t_{\rm acc}$ for \plc\ to accrete its GCR of $\sim1/6.7$. Grayed-out regions exclude scenarios whereby \plc\ has $t_{\rm acc} < 0.1$~Myr, which would require fine-tuning to prevent it from undergoing runaway gas accretion. The allowed regions of parameter space (green) suggest that \plc\ accreted material with a higher metallicity than the host star. The metallicity of the material accreted by \plb\ is likely lower than that of the host star, as the higher metallicity region is unrealistic.}
    \label{fig:dust_diff}
\end{figure}

Assuming both planets reached these GCRs in the same amount of time and that the temperature and the adiabatic index at their RCBs did not differ, we arrive at the following expression from \cite{Lee2021}:
\begin{equation}
    \bigg(\frac{\Sigma_{\rm b}}{\Sigma_{\rm c}}\bigg)^{0.12} \bigg(\frac{Z_{\rm c}}{Z_{\rm b}}\bigg)^{0.4} \bigg(\frac{\mu_{\rm b}}{\mu_{\rm c}}\bigg)^{3.4} \gtrsim \frac{1/2}{1/6.7},
\end{equation}
where $\Sigma$ is the gas surface density, $Z$ is the dust-to-gas ratio or equivalently the dust content of the accreted material, $\mu$ is the mean molecular weight, and the subscripts correspond to the two planets. For $\Sigma \propto r^{-1}$, the ratio $\Sigma_{\rm b} / \Sigma_{\rm c} = a_{\rm c} / a_{\rm b}$, i.e. the inverse ratio of their semi-major axes. Since $1 / \mu = Z / \mu_Z + (1-Z) / \mu_{H, He}$ with $\mu_Z = 17$ (approximate mean molecular weight of metals in a solar metallicity gas at the relevant temperature and pressure, \citealt{Woitke2018}) and $\mu_{H, He} = 2.32$ \citep{Asplund2021, Lodders2021}, this expression provides us with a preliminary estimate of the difference in $Z$ required for \plb\ to undergo runaway gas accretion while \plc\ accretes a modest envelope:
\begin{equation}
    \bigg(\frac{Z_{\rm c}}{Z_{\rm b}}\bigg)^{0.4} \, \bigg(\frac{1 - Z_{\rm c} + (\mu_{H, He} / \mu_Z) Z_{\rm c}}{1 - Z_{\rm b} + (\mu_{H, He} / \mu_Z) Z_{\rm b}}\bigg)^{3.4} \gtrsim \frac{1/2}{1/6.7} \, \bigg(\frac{a_{\rm b}}{a_{\rm c}}\bigg)^{0.12}
\end{equation}
The dust content that leads to the different formation outcomes for the \host\ planets is marked in color in Figure~\ref{fig:dust_diff}. We estimate the time required for \plc\ to reach a GCR of $1/6.7$ assuming $\Sigma_{\rm c} = 2000$ g cm$^{-3}$, adiabatic index of 0.17 and temperature of 2500 K at the RCB, and the planet's effective accretion radius $f_R = 0.2$ times its Hill radius \citep{Lee2021}:
\begin{multline}
    \bigg(\frac{t_{\rm acc, c}}{1 {\rm Myr}}\bigg)^{0.4} = \frac{{\rm GCR_c}}{0.06} f_R^{-1}  \bigg(\frac{Z}{0.02}\bigg)^{0.4}  \bigg(\frac{2.37}{\mu}\bigg)^{3.4} \\ 
    \bigg(\frac{5 M_{\oplus}}{M_{\rm core}}\bigg)^{1.8} \bigg(\frac{{\rm 2000 \, g \, cm^{-2}}}{\Sigma_{\rm c}} \bigg)^{0.12}
\end{multline}
We grey out the parameter space in which \plc\ accretes its envelope in less than 0.1~Myr to alleviate the issue of fine tuning the gas dissipation timescale. Figure~\ref{fig:dust_diff} shows that for \plc, the dust content of the material accreted must be above the stellar value to sufficiently suppress gas accretion. For \plb, the metallicity must either be much lower to speed up envelope cooling and accretion or so much higher that the high mean molecular weight of the accreted material expedites gas accretion. The high-$Z_p$ region for \plb\ can likely be ruled out because it is unrealistically high ($>0.5$).

The combination of high-$Z$ for the inner \plc\ and low-$Z$ for the outer \plb\ may be achieved if the two planets were separated by the water snowline \citep{Chachan2021}. This scenario has the caveat that the fragmentation velocities of ice-free grains need to be smaller than those of the icy grains, such that the smaller ice-free grains slow their radial drift and pile up in the inner disk, boosting the local $Z$ there. However, latest laboratory measurements suggest the material strength between these two types of grains may be more similar than previously thought \citep[e.g.,][]{Gundlach2018, Musiolik2019, Kimura2020}. Alternative mechanisms to alter the envelope opacity would require enhanced dust accretion onto \plc\ due to the presence of disk substructures \citep[e.g.,][]{Chen2020} or grain growth and expedited cooling in \plb's atmosphere \citep[e.g.,][]{Ormel2014, Piso2015} to explain the different gas accretion histories for \plb\ and c.

We briefly note that meridional flows that advect disk material within a planet's Hill sphere could limit its envelope's ability to cool and grow \citep{Ormel2015, Ali-Dib2020, Moldenhauer2021}. Studies that employ more realistic opacities and equations of state show that the effect of this recycling on envelope accretion is not as potent as initially supposed \citep{Zhu2021, Bailey2023, Savignac2023}. The effect of recycling is also distance-dependent: in hotter regions of the disk close-in to the star, it can lead to a slowdown in gas accretion but the exact magnitude of the effect depends on the inner disk conditions \citep{Bailey2023, Savignac2023}. However, although these uncertainties might affect the amount of gas accreted by super-Earth mass cores, they are unlikely to prevent a $\sim 26 \, M_{\oplus}$ from accreting substantially more gas than \plc\ possesses.

\subsection{Kepler-511 c's formation by giant impacts of sub-Neptunes}
\label{sec:mergers}

Since the two \host\ planets have puzzling different accretion histories, we explore if this is an outcome of \plc\ forming by mergers of two or three planets (sub-Neptunes) after the gas disk's dispersal. It is more plausible that lower mass sub-Neptunes, rather than a 30~$M_{\oplus}$ core, would accrete envelopes that are only $\sim 10\%$ of their mass. If these sub-Neptunes subsequently undergo a dynamical instability that leads to their coalescence into a planet such as \plc, we could explain why \plb\ turned into a gas giant while \plc\ did not. This is similar to the hypothesis of late-stage formation of super-Earths \citep{Lee2016, Dawson2016}, except we primarily consider dynamical instabilities after gas dissipation when damping by the gas does not act to stabilize the system. Dynamical instability after the gas disk dispersal is in line with the results of \cite{Dai2024}, who show that young ($< 100$ Myr) multiplanet systems are almost always found to be resonant and that these resonances get disrupted over $\sim 100$ Myr timescale to produce the observed orbital period ratios of mature planetary systems. Such an explanation would also have the advantage of forming \plc\ from inner planets that are akin to those observed around Sun-like stars, especially those interior to cold-Jupiters \citep{Zhu2018, Bryan2019}.  Given the age of the \host\ system ($7.5^{+1.7}_{-1.5}$ Gyr), it is plausible that a typical inner system of sub-Neptunes became unstable and merged to form \plc. An instability might arise if the eccentricities and/or mutual inclinations of the sub-Neptunes are excited to large enough values to lead to orbit crossing. These excitations may be a result of the planets' mutual interactions with each other or they may be driven by the outer \plb\ \citep[e.g.,][]{Pu2018, Poon2020}.

\begin{figure}
    \centering
    \includegraphics[width=\linewidth]{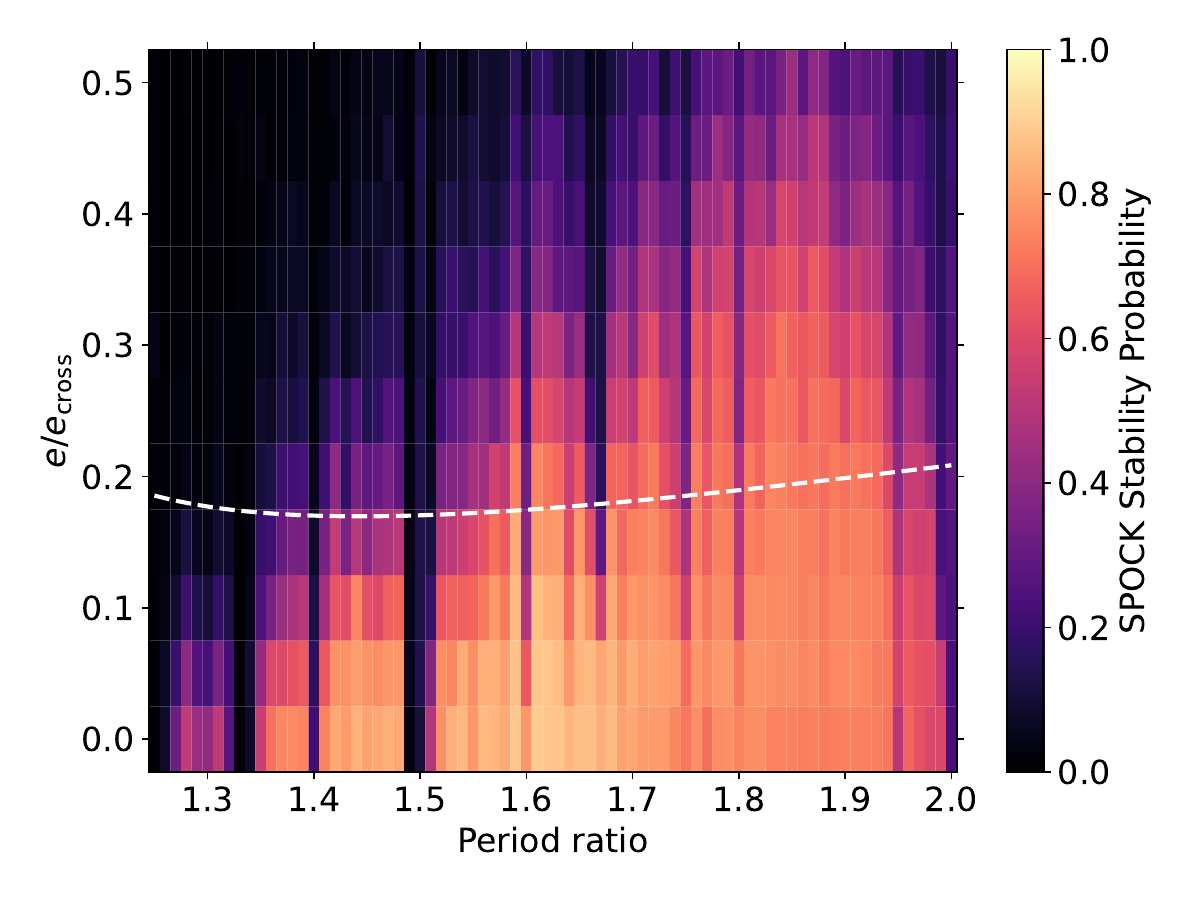}
    \caption{The stability probability for a system of three 10.5 M$_{\oplus}$ planets with a given eccentricity (as a fraction of the crossing eccentricity) and period ratio accompanied by an outer giant with the properties of \plb\ calculated by \textsf{SPOCK}. The dashed line indicates the rms time-averaged eccentricity of the three planets as a result of their interaction with the outer giant calculated via secular analytical theory.}
    \label{fig:spock_analysis}
\end{figure}

What causes instability in a tightly-packed system of planets is an area of ongoing research. In general, the onset of chaos is attributed to the overlap of two-body mean-motion resonances and three-body resonances \citep{Wisdom1980, Hadden2018, Petit2020, Lammers2024}. The time required for a tightly-packed chain of sub-Neptunes to undergo mergers is exponentially sensitive to the orbital spacing between the planets, their orbital eccentricities, and their mutual inclinations \citep{Pu2015}. Since the mergers are typically pairwise, it is difficult to conceive of more than two or three planets all merging together to form \plc. Therefore, we restrict our discussion to a system of three $10.5\,M_{\oplus}$ sub-Neptunes. The extent to which \plb\ can excite the inner planets' eccentricities and/or inclinations depends on the strength of their dynamical coupling, quantified by $\bar{\epsilon}$ as the ratio of the differential precession frequency of the inner planets with the outer giant and the mutual precession frequencies of the inner planets \citep{Pu2018}. We find that for period ratios $\mathcal{P}$ in the range of $1.2 - 2$ for the inner sub-Neptunes, $\bar{\epsilon}$ varies from $10^{-3} - 10^{-1}$. This implies that the inner sub-Neptunes would be more strongly coupled to each other than the outer giant and the giant planet alone is unlikely to drive the system to instability ($\bar{\epsilon} \gtrsim 1$ requires $\mathcal{P} \gtrsim 3.5$). Nonetheless, interactions with the outer giant planet can impart non-zero eccentricity to the inner sub-Neptunes and we use secular perturbation theory to calculate it \citep{Pu2018}. The rms time-averaged eccentricity of the three sub-Neptunes normalized by the crossing eccentricity ($e_{\rm cross} = (\mathcal{P}^{2/3} - 1) / (\mathcal{P}^{2/3} + 1)$) is shown in dashed white line as a function of the period ratio $\mathcal{P}$ of the three sub-Neptunes in Figure~\ref{fig:spock_analysis}.

We use \textsf{SPOCK} to determine the probability that a system of three sub-Neptunes with an outer giant planet is stable over timescales of $10^9$ orbits \citep{Tomayo2020}. \textsf{SPOCK} uses \textsf{REBOUND} \citep{Rein2012} to integrate orbits of an input planet configuration over the first $10^4$ orbits and measures summary statistics that are then used to predict stability probability from a machine-learning model trained on a set of $\sim$100,000 scale-invariant simulations of 3-planet systems. \cite{Tomayo2020} and \cite{Sobski2023} have shown that \textsf{SPOCK} generalizes well to planet configurations it was not trained on and that it agrees with prior stability metrics. We therefore adopt it to evaluate whether \plc\ could have formed from mergers of a system of inner sub-Neptunes.

Period ratios $\mathcal{P}$ of the inner sub-Neptunes are incremented by 0.01 in the range $1.25 - 2$ and their eccentricities are set as a fraction of the orbit-crossing eccentricity (in increments of 0.05 in the range $0 - 0.5$). For each combination of $\mathcal{P}$ and $e$, we calculate the stability probability for 100 different realizations of the arguments of pericenter and the initial mean anomalies (each randomly sampled from a $\mathcal{U}[0, 2 \pi)$ and average them to obtain a mean estimate that is shown by color in Figure~\ref{fig:spock_analysis}. The stability probability is lowest for $\mathcal{P}$ that correspond to integer ratios, as expected from the near-resonant interactions of the inner sub-Neptunes. Higher eccentricities also lower the probability that the system is stable over long timescales. Eccentricity pumping due to secular interactions with \plb\ therefore makes mergers of the inner sub-Neptunes more likely. In general, a larger part of the parameter space is stable at larger $\mathcal{P}$ but this trend would likely reverse when $\bar{\epsilon} \gtrsim 1$ ($\mathcal{P} \gtrsim 3.5$) and the influence of the giant planet becomes substantial.

If the sub-Neptune progenitors were close to a mean-motion resonance, they would easily undergo the mergers necessary to produce a planet such as Kepler-511 c. For $\mathcal{P}$ not close to mean-motion resonances, the stability probability can be as high as $0.4 - 0.8$ even when planet eccentricities are set by secular perturbation from \plb. We note that $10^9$ orbits of the innermost planet only correspond to $\sim 70$ Myr, which is a mere 1\% of the system age. These stability probabilities should be treated as upper limits and they could be much lower over Gyr timescales ($10^{11}$ orbits). Indeed, the median and 84th percentile instability times for orbital configurations with stability probability $>0.4$ (when $e$ is set by \plb, dashed line in Figure~\ref{fig:spock_analysis}) are lower than the age of the system \citep{Cranmer2021}. Mergers therefore provide a plausible albeit a low likelihood pathway for the formation of \plc\ like planets in some planetary systems, which would be commensurate with their low occurrence rates \citep[see also][for a similar mechanism to produce Neptune-like planets around very low mass stars]{Liveoak2024}.

The process by which \plb\ came to acquire its significant eccentricity also might have played a role in destabilizing any inner planets that could coalesce to form \plc. For example, if the outer unconfirmed companion scattered \plb\ to its current orbit, such an event could have destabilized an inner system of sub-Neptunes. A more precise measurement of \plc's eccentricity (our current estimate is consistent with zero) could provide more insight into its dynamical history and the plausibility of mergers as an origin channel for planets like \plc.

The plausibility of this hypothesis also depends on the impact of mergers on the atmosphere retention of the final planet. The outcome of the merger on atmosphere loss depends on the mass ratio, impact velocity, and the geometry of the impact as well as the post-merger thermal evolution \citep{Asphaug2010, Leinhardt2012, Stewart2014, Inamdar2016, Schlichting2018, Biersteker2019}. Since the probability of having an impact parameter $\leq b$ scales as $b^2$, planets are much more likely to collide at oblique angles rather than head-on. In our scenario with $\sim 10$~M$_\oplus$ bodies, impact velocities in the range $\sim e_{\rm cross} v_{\rm K} - v_{\rm esc}$ ($v_{\rm K}$ and $v_{\rm esc}$ are the Keplerian and mutual escape velocities, see also \citealt{Liveoak2024} who find sub-escape impact velocities), and the higher likelihood of an oblique impact suggests that mergers will likely strip a small fraction of the atmosphere \citep{Denman2020, Denman2022, Naponiello2023}. We encourage continued exploration of the impact of mergers on envelope retention for the tightly-packed close-in exoplanets.


\section{Summary}
\label{sec:conc}

In this work, we collected RV measurements of \host\ over an 8.5-year baseline (Figure~\ref{fig:rv}) to dynamically confirm its two (previously validated) transiting exoplanets. \plb\ is a gas giant planet with $M_p = 0.44^{+0.11}_{-0.12}$~$M_{\rm J}$ on a 297~day, moderately eccentric orbit. Given its unlikely transiting orbit, \plb\ is a prime member of the Giant Outer Transiting Exoplanet Mass (GOT `EM) survey \citep{Dalba2021a, Dalba2021c, Mann2023}. Its inner ($P=27$~days) companion (\plc) has a mass of $32^{+11}_{-12}$~$M_{\rm earth}$ and is three time as dense. We also identify a long-term trend in the RV time series data that is most likely indicative of an additional massive object (planetary or otherwise) in the outer reaches of the \host\ system (Section~\ref{sec:companion}).

We take advantage of both planets' relatively long orbits and low stellar irradiation (compared to other transiting exoplanets) to infer their bulk metallicities (Section~\ref{sec:zp}). Interestingly, both planets have similar masses of heavy elements ($31^{+12}_{-10}$~$M_{\oplus}$ and $26^{+11}_{-10}$~$M_{\oplus}$ for planets b and c respectively) but only \plb\ seems to have accreted a massive gaseous envelope. Since the ability of planets to accrete primordial gas depends on their heavy element content, the \host\ planets are an intriguing duo that pose an informative challenge to planet formation and evolution theories.

We explore three scenarios that could account for the planets' current properties. Firstly, we rule out envelope mass loss as a means to explaining the difference in bulk metallicities of these planets: at 27 days, \plc\ is simply too far away from its host star to suffer significant mass loss. Secondly, we evaluate a variety of reasons why their gas accretion histories might have differed. Although differences in the dust content of the disk gas accreted by these planets could drive their divergent accretion histories, it is difficult to attain such strong contrasts in the local dust content. Finally, we consider the scenario that \plc\ is gas-poor because it formed from the merger of lower mass cores (that accrete less gas) after the dissipation of the protoplanetary disk. We show that secular perturbations from \plb\ could pump up the eccentricities of such an inner system of sub-Neptunes and render their merger more likely over long timescales. 

Overall, the \host\ system serves as an important benchmark for numerous reasons. It contains multiple giant planets; both planets transit, enabling a measurement of their radii; the outer planet has an exceptionally long orbital period considering its transiting geometry, leaving it with an irradiation only 4.5 times that of the Earth and an equilibrium temperature of $\sim$400~K; and the two planets seem to have divergent accretion histories. We encourage additional theoretical explorations to provide an explanation for how systems of multiple giant planets like \host\ could have formed.  


\section*{acknowledgments}
The authors recognize the cultural significance and sanctity that the summit of Maunakea has within the indigenous Hawaiian community. We are deeply grateful to have the opportunity to conduct observations from this mountain. We acknowledge the impact of our presence there and the ongoing efforts to preserve this special place.

The authors thank all of the observers in the California Planet Search team for their many hours of hard work in collecting the RVs published here. We thank an anonymous referee for their constructive comments, especially regarding the nature of the host star, that led us to sharpen our analysis. The authors are grateful to Jonathan Fortney, Sarah Millholland, and Yanqin Wu for helpful conversations about this planetary system. We are particularly indebted to Adam Kraus for an enlightening and extremely useful exchange on the possible binarity of this system. Part of this research was conducted at the Other Worlds Laboratory Summer Program 2022 at UC Santa Cruz, which is generously supported by the Heising-Simons Foundation. P.D. acknowledges support by a 51 Pegasi b Postdoctoral Fellowship from the Heising-Simons Foundation and by a National Science Foundation (NSF) Astronomy and Astrophysics Postdoctoral Fellowship under award AST-1903811. M.P. gratefully acknowledges NASA award 80NSSC22M0024.

This research has made use of the NASA Exoplanet Archive, which is operated by the California Institute of Technology, under contract with the National Aeronautics and Space Administration under the Exoplanet Exploration Program. This paper includes data collected by the \kepler\ mission and obtained from the MAST data archive at the Space Telescope Science Institute (STScI). Funding for the Kepler mission is provided by the NASA Science Mission Directorate. STScI is operated by the Association of Universities for Research in Astronomy, Inc., under NASA contract NAS 5–26555. This research has made use of the Exoplanet Follow-up Observation Program (ExoFOP; DOI: 10.26134/ExoFOP5) website, which is operated by the California Institute of Technology, under contract with the National Aeronautics and Space Administration under the Exoplanet Exploration Program.

Some of the data presented herein were obtained at the W. M. Keck Observatory, which is operated as a scientific partnership among the California Institute of Technology, the University of California, and NASA. The Observatory was made possible by the generous financial support of the W. M. Keck Foundation. Some of the Keck data were obtained under PI Data awards 2013A and 2013B (M. Payne). 


\vspace{2mm}
\facilities{Keck:I (HIRES), Kepler}\\

\vspace{2mm}
\software{   \textsf{astropy} \citep{astropy2013,astropy2018},
                \textsf{corner} \citep{ForemanMackey2016a},
                \textsf{EXOFASTv2} \citep{Eastman2013,Eastman2017,Eastman2019}, 
                \textsf{lightkurve} \citep{Lightkurve2018},
                \textsf{RadVel} \citep{Fulton2018}, 
                \textsf{SpecMatch}, \citep{Petigura2015,Petigura2017b},
                \textsf{pymc3} \citep{pymc3},
                \textsf{theano} \citep{theano},
                \textsf{REBOUND} \citep{Rein2012},
                \textsf{SPOCK} \citep{Tomayo2020},
                \textsf{exoplanet} \citep{Agol2020,Kipping2013b,exoplanet:joss,exoplanet:zenodo},
                \textsf{celerite2} \citep{ForemanMackey2017,ForemanMackey2018}, 
                \textsf{starry} \citep{Luger2018}, 
                \textsf{arviz} \citep{arviz}}


\appendix

\section{Limits on the presence of a companion from RV spectra}
\label{sec:kolbl_analysis}

\begin{figure*}
    \includegraphics[width=0.5\linewidth]{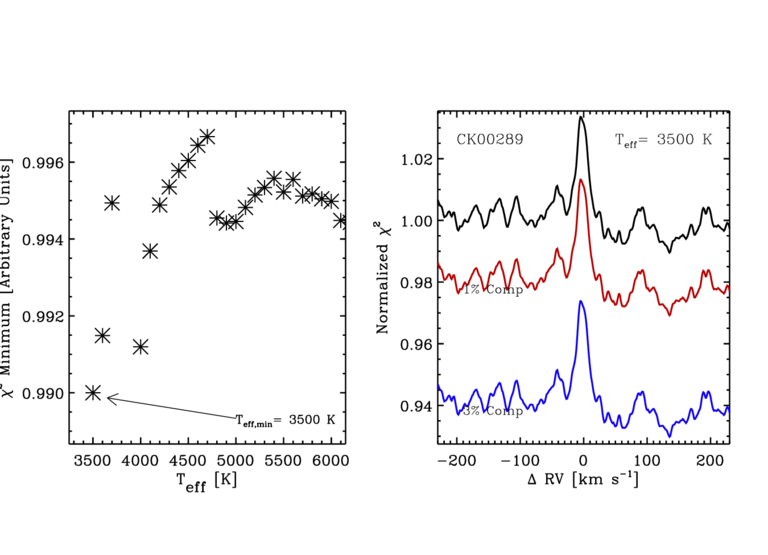}
    \includegraphics[width=0.5\linewidth]{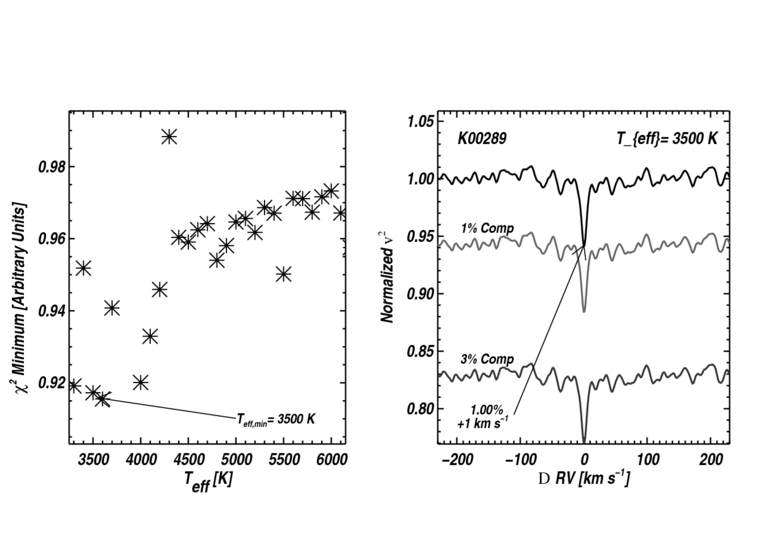}
    \caption{RealMatch analysis plots for two of our RV spectra \citep{Kolbl2015}. We rule out any secondary companions brighter than 1\% the primary's brightness with RV separation of $\geq \pm 10$ km s$^{-1}$.}
    \label{fig:kolbl_analysis}
\end{figure*}

We performed the RealMatch analysis laid out in \cite{Kolbl2015} on our RV spectra for \host. The results of this analysis for two of our spectra are shown in Figure~\ref{fig:kolbl_analysis}. We are sensitive to companions that are 1\% the brightness of the primary and $\Delta$RV separations of $\geq \pm 10$ km s$^{-1}$. Within these limits, we do not detect any secondary companions. Although this analysis does not eliminate a nearly face-on equal mass companion directly, we present additional arguments that disfavor such a companion in \S~\ref{sec:stellar_prop}.

\section{Fits with and without stellar density constraints}
\label{sec:stellar_density}

\begin{figure*}
    \includegraphics[width=\linewidth]{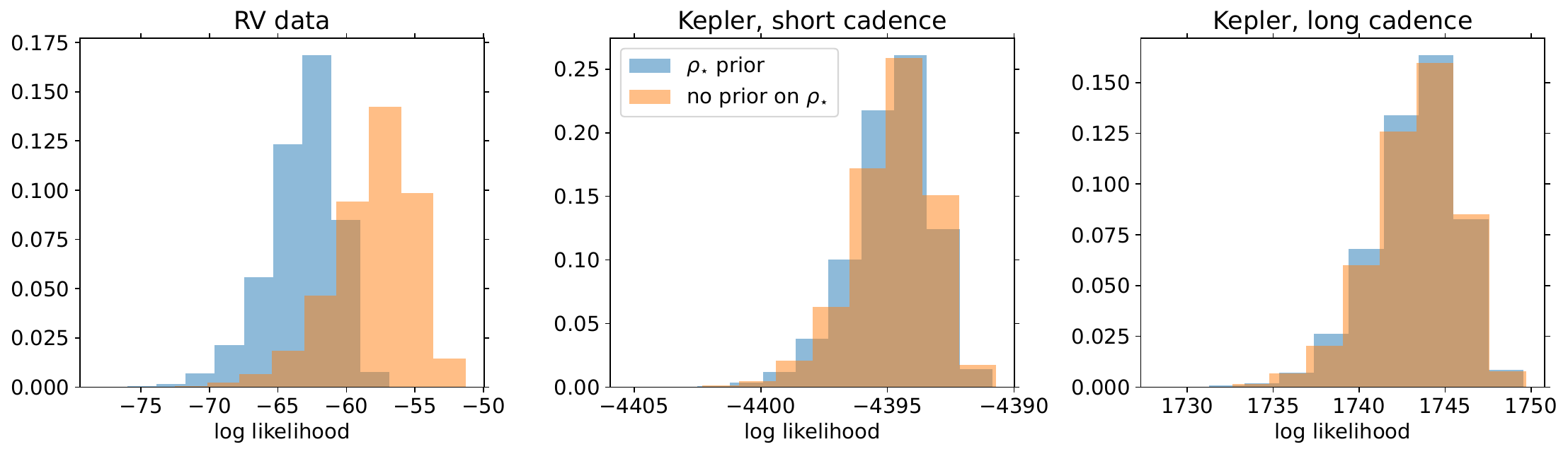}
    \caption{The histograms for log likelihoods of our fits to the RV data and short cadence and long cadence transit data. The preference for a fit without a prior on the stellar density is driven by the RV data, which bias the fitted eccentricity and argument of periastron of \plb.}
    \label{fig:stellar_rho_loglike}
\end{figure*}

\begin{figure}
    \centering
    \includegraphics[width=0.49\linewidth]{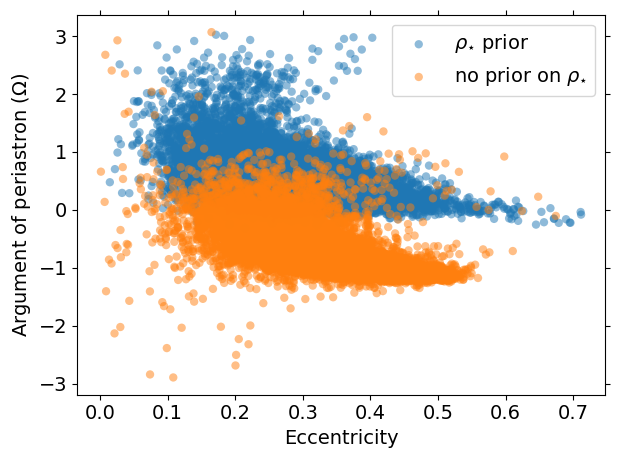}
    \includegraphics[width=0.49\linewidth]{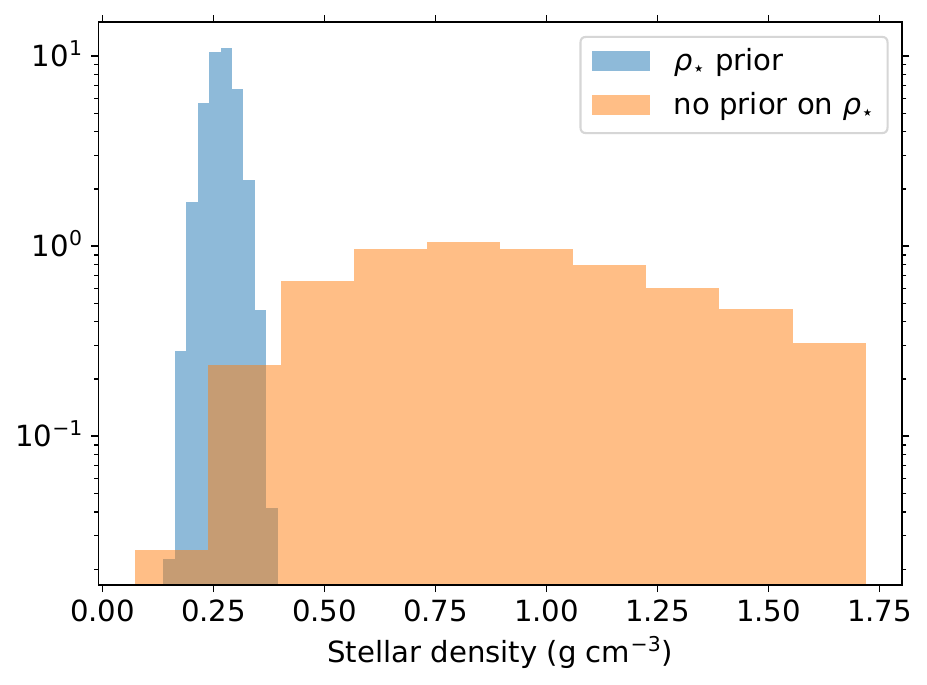}
    \caption{The left panel shows the fitted eccentricity and argument of periastron of \plb\ with and without a prior on the stellar density. The right panel compares our stellar density prior with its fitted posterior when no prior is placed.}
    \label{fig:stellar-rho-prior-comparison}
\end{figure}

Figure~\ref{fig:stellar_rho_loglike} shows the difference in log likelihood of the fits to the RV and transit data with and without a prior on the stellar density. The higher log likelihood for the fit without a prior on the stellar density is driven entirely by the RV data. The sparsely sampled RV data bias the fitted eccentricity and argument of periastron, which in turn affects the fitted $a/R_\star$ from the transit data and the resulting stellar density (Figure~\ref{fig:stellar-rho-prior-comparison}). To mitigate this effect, \cite{Espinoza2019} recommend placing a prior on the stellar density instead. For planets with high eccentricities, this choice can also be used to better constrain their eccentricities via the `photoeccentricity' effect \citep{Dawson2012a}.

\bibliography{references}{}
\bibliographystyle{aasjournal}

\end{document}

%% file: planet_table_RHO_STAR.tex
\begin{deluxetable*}{llcc}
\tabletypesize{\scriptsize}
\tablecaption{Median Values and 68\% Confidence Interval for the \host\ Planet Parameters}
\tablehead{\colhead{Parameter} & \colhead{Units} & \multicolumn{2}{c}{Values}}
\startdata
\smallskip\\\multicolumn{2}{l}{}&\plb&\plc\smallskip\\
~~~~$P$ &Period (days) & $296.63766\pm0.00039$ & $26.629424\pm0.000039$\\
~~~~$R_p$ &Radius (\rj) & $0.886\pm0.033$ & $0.371^{+0.018}_{-0.017}$ \\
~~~~$R_p$ &Radius (\re) & $9.93\pm0.37$ & $4.16^{+0.20}_{-0.19}$ \\
~~~~$M_p$ &Mass (\mj) & $0.44^{+0.11}_{-0.12}$ & $0.100^{+0.036}_{-0.039}$ \\
~~~~$M_p$ &Mass (\me) & $140^{+35}_{-38}$ & $32^{+11}_{-12}$ \\
~~~~$T_C$ &Time of Conjunction (\bjdtdb) & $2455236.67214^{+0.00083}_{-0.00082}$ & $2454971.7374\pm0.0013$ \\
~~~~$a$ &Semi-major Axis (au) & $0.894\pm0.048$ & $0.1792\pm0.0097$ \\
~~~~$i$ &Inclination (degrees) & $89.719^{+0.029}_{-0.037}$ & $88.59^{+0.38}_{-0.75}$ \\
~~~~$e$ &Eccentricity  & $0.277^{+0.070}_{-0.068}$ & $0.17^{+0.14}_{-0.11}$ \\
~~~~$\omega_*$ &Argument of Periastron (degrees) & $26^{+19}_{-13}$ & $18^{+89}_{-95}$ \\
~~~~$T_{eq}$ &Equilibrium Temperature$^{1}$ (K) & $402\pm14$ & $899\pm32$ \\
~~~~$K$ &RV Semi-amplitude (m~s$^{-1}$) & $13.7^{+3.4}_{-3.8}$ & $6.8^{+2.4}_{-2.6}$ \\
~~~~$R_p/R_*$ &Radius of Planet in Stellar Radii  & $0.04986^{+0.00054}_{-0.00056}$ & $0.02089^{+0.00066}_{-0.00061}$ \\
~~~~$a/R_*$ &Semi-major Axis in Stellar Radii  & $107.6\pm4.4$ & $21.57\pm0.89$\\
~~~~$\delta$ & Transit Depth $\left(R_p/R_*\right)^2$  & $0.002486^{+0.000054}_{-0.000056}$ & $0.000436^{+0.000028}_{-0.000025}$ \\
~~~~$b$ &Transit Impact Parameter  & $0.527^{+0.049}_{-0.066}$ & $0.53^{+0.14}_{-0.28}$ \\
~~~~$\rho_p$ &Density (g~cm$^{-3}$) & $0.84^{+0.23}_{-0.25}$ & $2.6^{+1.0}_{-1.1}$ \\
~~~~$S_p$ & Insolation ($S_{\earth}$) & $4.37\pm0.54$ & $109\pm13$ \\
\smallskip\\\multicolumn{2}{l}{Kepler Limb-darkening Parameters:}& \smallskip\\
~~~~$u_{1}$ &Linear Coefficient  & \multicolumn{2}{c}{$0.524^{+0.097}_{-0.085}$}  \\
~~~~$u_{2}$ &Quadratic Coefficient  & \multicolumn{2}{c}{$0.02^{+0.13}_{-0.14}$} \\
\smallskip\\\multicolumn{2}{l}{Keck-HIRES Parameters:}& \smallskip\\
~~~~$\gamma$ &Relative RV Offset$^{2}$ (m~s$^{-1}$) & \multicolumn{2}{c}{$14.13^{+0.94}_{-0.96}$} \\
~~~~$\dot\gamma$ & RV Linear Trend Coefficient$^{2}$ (m~s$^{-1}$~d$^{-1}$) & \multicolumn{2}{c}{$-0.0246^{+0.0015}_{-0.0014}$} \\
~~~~$\ddot\gamma$ & RV Quadratic Trend Coefficient$^{2}$ (m~s$^{-1}$~d$^{-2}$) & \multicolumn{2}{c}{$-0.0000047^{+0.0000012}_{-0.0000013}$} \\
~~~~$\sigma_J$ &RV Jitter (m~s$^{-1}$) & \multicolumn{2}{c}{$5.1^{+1.8}_{-1.4}$} \\
\smallskip\\\multicolumn{2}{l}{Kepler Nuisance Parameters:}& Short Cadence & Long Cadence \smallskip\\
~~~~$\sigma_F^{2}$ & Added Flux Variance (ppm)  & $465.4^{+4.1}_{-4.0}$ & $95.6\pm1.6$ \\
~~~~$F_0$ &Baseline Flux  & $1.0000\pm0.0060$ & $0.9999\pm0.0029$ \\
\enddata
\label{tab:planet}
\tablenotetext{1}{Assumes no albedo and perfect redistribution.}
\tablenotetext{2}{Relative to time BJD$_{\rm TDB}$ = 2457998.3347095.}
\end{deluxetable*}